\documentclass[%
 aip,
 jmp,%
 amsmath, amssymb,
 reprint,%
]{revtex4-2}
\usepackage[top=1in, bottom=1in, left=1in, right=1in]{geometry}

\usepackage{xcolor}
\usepackage[utf8]{inputenc}
\usepackage[T1]{fontenc}
\usepackage{textcomp}
\usepackage{graphicx}% Include figure files
\usepackage{dcolumn}% Align table columns on decimal point
\usepackage{bm}% bold math
\usepackage{hyperref}
\usepackage{ulem}
\begin{document}

% \preprint{AIP/123-QED}

\title{Attosecond Control of Squeezed Light}

%\title{Strong-field Driven Subcycle Band Modulation Measured with Electric Field Observables in Four-wave Mixing}

 \author{Russell Zimmerman$^{\dagger}$}
 % \thanks{These authors contributed equally to this work.}
 \affiliation{Department of Physics and Astronomy, Purdue University, 525 Northwestern Ave., West Lafayette, Indiana 47907, USA}
 
\author{Shashank Kumar$^{\dagger}$}%
\email{kumar414@purdue.edu}
 % \thanks{These authors contributed equally to this work.}
 \affiliation{Department of Physics and Astronomy, Purdue University, 525 Northwestern Ave., West Lafayette, Indiana 47907, USA}

\author{Shiva Kant Tiwari}
\affiliation{Department of Chemistry,  Purdue University, 560 Oval Dr, West Lafayette, Indiana 47907, USA}

\author{Md Ehsanuzzaman}
  \affiliation{Department of Physics and Astronomy, Purdue University, 525 Northwestern Ave., West Lafayette, Indiana 47907, USA}

\author{Eric Liu}
  \affiliation{Department of Physics and Astronomy, Purdue University, 525 Northwestern Ave., West Lafayette, Indiana 47907, USA}

 \author{Francis Walz}
  \affiliation{Department of Physics and Astronomy, Purdue University, 525 Northwestern Ave., West Lafayette, Indiana 47907, USA}

\author{Siddhant Pandey}
 \affiliation{Department of Physics and Astronomy, Purdue University, 525 Northwestern Ave., West Lafayette, Indiana 47907, USA}

\author{George J. Economou II}
  \affiliation{Department of Physics and Astronomy, Purdue University, 525 Northwestern Ave., West Lafayette, Indiana 47907, USA}

\author{Hadiseh Alaeian}
\affiliation{Elmore Family School of Electrical and Computer Engineering, Purdue University, 465 Northwestern Ave., West Lafayette, Indiana 47907, USA}

\affiliation{Department of Physics and Astronomy, Purdue University, 525 Northwestern Ave., West Lafayette, Indiana 47907, USA}
 \affiliation{Purdue Quantum Science and Engineering Institute, 1205 W State St, West Lafayette, Indiana 47907, USA}

\author{Chen-Ting Liao}
\affiliation{Department of Physics, Indiana University Bloomington, 727 E. Third St., Bloomington, Indiana 47405, USA}

\author{Valentin Walther}
\affiliation{Department of Physics and Astronomy, Purdue University, 525 Northwestern Ave., West Lafayette, Indiana 47907, USA}
 \affiliation{Purdue Quantum Science and Engineering Institute, 1205 W State St, West Lafayette, Indiana 47907, USA}
 \affiliation{Department of Chemistry,  Purdue University, 560 Oval Dr, West Lafayette, Indiana 47907, USA}

 \author{Niranjan Shivaram}
 \email{niranjan@purdue.edu}
  \affiliation{Department of Physics and Astronomy, Purdue University, 525 Northwestern Ave., West Lafayette, Indiana 47907, USA}
 \affiliation{Purdue Quantum Science and Engineering Institute, 1205 W State St, West Lafayette, Indiana 47907, USA}

% \affil[3]{\orgdiv{Molecular Foundry}, \orgname{Lawrence Berkeley National Laboratory}, \orgaddress{\street{67 Cyclotron Rd.}, \city{Berkeley}, \postcode{94720}, \state{California}, \country{USA}}}

\begin{abstract}
        
    Squeezed light is a key resource in quantum metrology and quantum information science. It is primarily generated through nonlinear optical interactions, where the degree of squeezing is set by the nonlinearity of the medium. Here, we modulate the third-order nonlinear response of a dielectric with strong ultrafast laser fields to control squeezed light generation on attosecond time scales. By tuning a sub-cycle phase delay between the input femtosecond pulses, we switch the generated light between amplitude-squeezed and phase-squeezed states. We measure the quantum noise using a frequency-resolved balanced homodyne detection scheme that extracts field quadratures in many frequency modes simultaneously. From these measurements we obtain the complete coherency matrix containing quadrature correlations across the frequency modes of the pulse. These results enable quantum light sources with sub-cycle control of squeezing and open a route to transduction of dynamical quantum correlations in matter to quantum correlations in electric fields.
\end{abstract}

\maketitle

\noindent $^{\dagger}$These authors contributed equally to this work.

%In ultrafast AMO research squeezed light can be used to study quantum field observables previously unmeasurable by classical light sources\textbf{[CITATION?]}.

%In recent years, there have been numerous theoretical studies of quantum noise of a pulse propagating through a nonlinear waveguide \cite{Yanagimoto2021, Ng2023}. 

%Highly pure optical Schr{\"o}dinger's cat states have found usage in quantum information processing.

%Experimental benchmarks in pulsed laser quantum state detection have only occurred more recently\cite{wenger_pulsed_2004}.

%experimentally describes the degree of squeezing of an ultrafast laser pulse due to a modulation of nonlinear response, signifying the first instance of attosecond control of quantum squeezing. This is achieved by attuning the relative phase of input ultrafast beams within a degenerate four-wave mixing (DFWM) interaction. The DFWM signal is then measured using an optical homodyne detection scheme. We see that our measurement is near the shot-noise limit and show control of squeezing ranging up to 2 dB\textbf{[CHECK THIS WITH BEST DATA SET]}. The time steps between the input pulses that drive the non-linear process are as precise as 27 attoseconds.

\section*{Introduction}\label{sec1}
\begin{figure}[t]
    \centering
    \includegraphics[width=\linewidth]{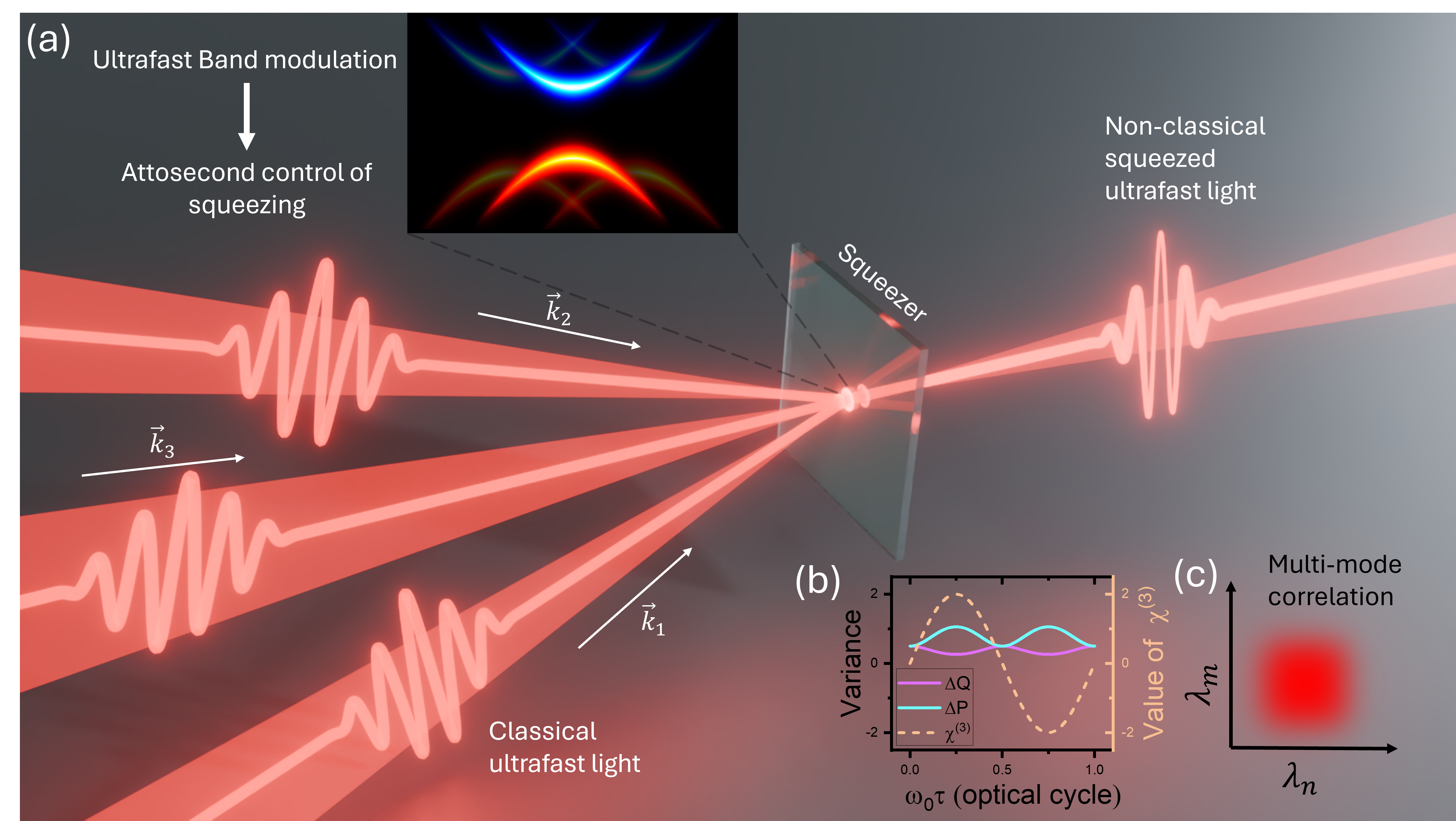}
    \caption{Concept and Experimental Scheme: (a) Three femtosecond pulses interact with a large band-gap dielectric (MgO) and generate a four-wave mixing signal exhibiting quadrature squeezing. The four-wave mixing process occurs simultaneously with a modulation of the material band-structure (inset in (a)) driven by the strong electric field of the laser pulses. This band modulation results in attosecond scale control of the degree and type of quadrature squeezing. The quantum state of this broadband squeezed light is measured with a frequency-resolved homodyne tomography technique revealing multi-frequency mode quadrature correlations in the generated ultrafast squeezed light. (b) shows quadrature variance modulation from theory and (c) a schematic of quadrature correlations between different frequency modes that can be extracted from the experiment. }
    \label{fig:concept}
\end{figure}
Squeezed light is a crucial component in a variety of quantum applications ranging from quantum metrology to quantum information science \cite{andersen2016}. At the Laser Interferometer Gravitational Wave Observatory (LIGO), squeezed light has been utilized to reduce shot noise below the quantum limit, which allows for unprecedented sensitivity of gravitational waves \cite{Aasi2013, PhysRevX.13.041021, Oelker:14, PhysRevLett.123.231108}. In the field of quantum computing, the relevance of squeezed light to continuous-variable quantum information processing allows a better understanding of dense coding and universal quantum computing \cite{RevModPhys.77.513, Lvovsky2015}. Exploiting the quantum states of light is also useful in imaging applications to reach resolution beyond the classical limit \cite{Moreau2019, Casacio2021}. Squeezed light with its quadrature noise below the quantum noise limit is fundamental in enhancing both laser interferometry \cite{Schnabel2017} as well as spectroscopy \cite{Polzik1992}. Despite achieving initial generation nearly four decades ago, squeezed light has primarily been generated with continuous-wave lasers \cite{Shelby1986,Wu98,Mehmet2011}. In addition to squeezed light, other quantum states of light have been generated, such as optical Schr{\"o}dinger's cat states, which are superpositions of two coherent states \cite{Asavanant2017,lewenstein2021}. These experiments use homodyne detection and have primarily remained within the continuous-wave laser regime, though recent experiments have measured these cat states from ultrafast lasers \cite{lewenstein2021}. 

While continuous-wave squeezed light sources have been available for a few decades, the generation and measurement of ultrafast squeezed light on the femtosecond and attosecond scale is still in its infancy. Bright squeezed vacuum states have been generated with ultrashort pulses \cite{Cutipa2022,Barakat2025} and, more recently, have been used to drive high harmonic generation in solids \cite{Rasputnyi2024} and gases \cite{Tzur2025}. Compared to continuous-wave lasers, ultrafast lasers present additional challenges, such as difficulty in spatial mode cleaning and decreased power stability of the laser. Alongside these challenges, ultrafast lasers also offer new opportunities for quantum light generation. The multimode structure of the generated quantum state is set by the joint action of the broadband driving fields and the spectrally broad nonlinear response of the medium through the phase-matched mixing process. In addition, the strong fields of ultrafast lasers can manipulate the nonlinear response of the material used to generate the quantum light. In the femtosecond regime, squeezed vacuum has been generated and measured in the time-domain on a sub-cycle scale using electro-optic sampling \cite{Riek2017}. Recently, few-cycle femtosecond pulses at visible and near infrared wavelengths were used to generate ultrafast light for which amplitude squeezing was reported \cite{Sennary2025}. In that work, the Wigner distributions and the degree of squeezing were inferred by comparison with theory. The reported squeezing was not referenced to an independently calibrated shot-noise level. Such a reference is essential in a demonstration of squeezing, since squeezing means quadrature noise below the shot-noise level. Furthermore, complete quantum state tomography of extreme-ultraviolet light from high harmonic generation has now been demonstrated \cite{tzur2025_fluctuation}.  

Nonlinear optical interactions typically result in the generation of quantum light \cite{Bachor2019, Drummond2014, Kwiat1995}. Using such interactions quantum light demonstrating quadrature squeezing has been generated via degenerate four-wave mixing (DFWM) \cite{kumar1984, Bondurant1984, SchmittbergerMarlow2020, Marhic1991} and second harmonic generation \cite{Vahlbruch2016}. Despite major advances in squeezed light generation, ultrafast nonlinear interactions that determine its transient non-classical properties  including the potential for attosecond dynamical control have not been explored. Achieving a high degree of control over the properties of squeezed light and the capabilities to measure quantum correlations on attosecond time scales could greatly benefit the emerging field of attosecond quantum optics. Here, we describe an experiment to generate, measure, and control ultrafast squeezed light from DFWM in magnesium oxide (see figure \ref{fig:concept}). We demonstrate sub-cycle, attosecond-scale control of both the degree of squeezing and the type of quadrature squeezing. The switching of squeezing between the amplitude and phase quadratures is prominently observed. This control over squeezed light generation is achieved using a strong-field-driven nonlinear response modulation (NRM) in the generating material \cite{walz2025}. We completely characterize the generated ultrafast squeezed light pulse using frequency-resolved balanced homodyne detection (FR-BHD). This FR-BHD is achieved using a modulation of the nonlinear response of the squeezing medium induced by varying the time delay between the input DFWM pulses. Such a strong-field-driven modulation of the nonlinear response modulates the generated broadband squeezed light and is used as a filter to suppress electronic noise and measure frequency-resolved squeezing below the shot-noise limit of the laser light. In contrast to balanced photodiode measurements, the filtering process demonstrated here (henceforth called NRM filtering) enables homodyne tomography measurements at the quantum noise limit with a CMOS camera, which has not been previously possible. Furthermore, our FR-BHD measurement provides access to quadrature correlations between different frequency modes measured simultaneously. This in turn allows direct extraction of the coherency matrix and helps us determine the principal modes contributing to our squeezed light pulse \cite{Fabre2020}. In previous work, frequency multiplexing has been utilized to measure correlations between different frequency modes in squeezed light generated in an optical parametric oscillator \cite{Roslund2013}. Our approach goes beyond such multiplexing techniques by offering a path to measure dynamical quadrature correlations to study ultrafast quantum dynamics.

\section*{Results and Discussion}
% Here we should include the following figures and corresponding discussion:

% \noindent1. Fig 1 will show the experimental setup and measurement scheme. This should include an actual image of a balanced spectrometer data showing both spectra.\\
% 2. 1g 2 will show a schematic of our two-dimensional filtering scheme and how we apply it to each {X,Y} quadrature value  the Wigner plot. I will share Sidd's Opt Exp paper that is in prep as an example to help with this visualization.\\
% 2. 1g 2 will show a schematic of our two-dimensional filtering scheme and how we apply it to each {X,Y} quadrature value  the Wigner plot. I will share Sidd's Opt Exp paper that is in prep as an example to help with this visualization.\\
% 3. Fig 3 will show multiple snapshots of the badpassed Wigner plot at different time delays.\\
% 4. Fig 4 will show the fitted ellipse amplitude and phase axis as a function of time delay and possible comparison with theoretical calculations of squeezing modulation.\\
% 5. Possibly include correlations between multiple wavlengths within the bandwidth.

\begin{figure}[t]
    \centering
    \includegraphics[width=\linewidth]{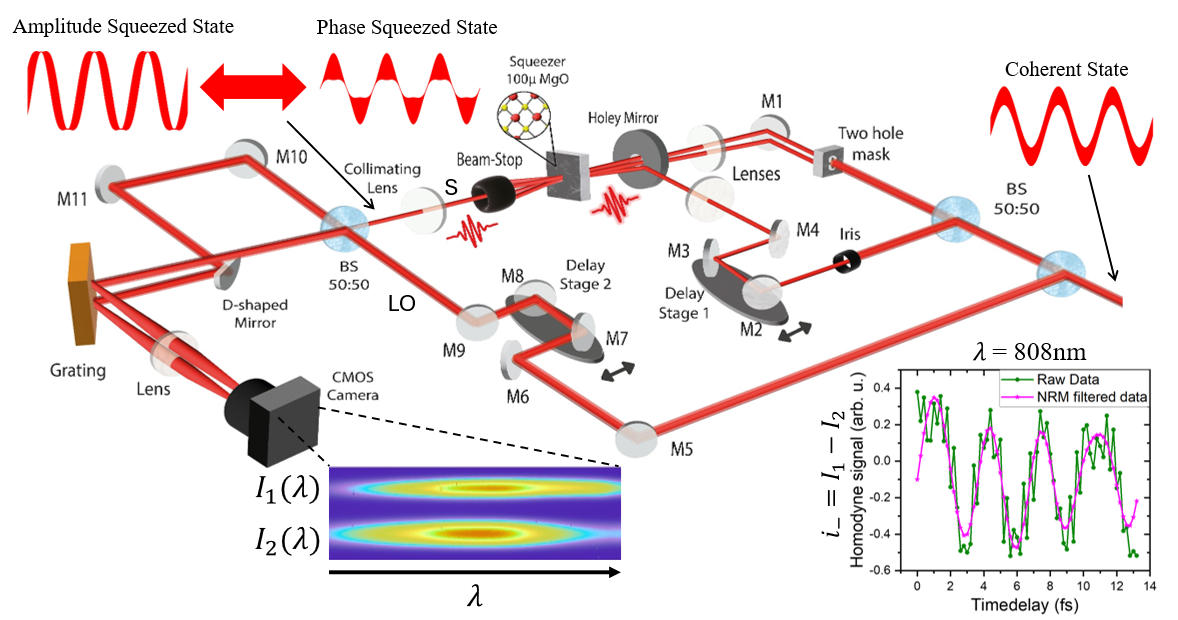}
    \caption{Schematic of ultrafast squeezed light generation with attosecond control and frequency-resolved balanced homodyne detection using a grating spectrometer. A 50-fs laser beam is split to form the local oscillator and a squeezed light generation beam. The squeezed light is generated using a degenerate four-wave mixing (DFWM) scheme using three beams created using a second beamsplitter and a two-hole mask. The DFWM interaction occurs in a 100 micron thick MgO target (squeezer) placed at the focus. The DFWM signal beam (squeezed light) is collimated using a lens and sent to the balanced homodyne setup, where it is combined with the local oscillator using a beamsplitter. The transmitted and reflected beams from the beamsplitter propagate to a diffraction grating, and the resulting spectra are imaged with a CMOS camera. Delay stage 1 introduces an attosecond-scale time delay ($\tau$) between the DFWM pulses for attosecond squeezing control. The phase delay ($\phi_{LO}$) between the squeezed light and the local oscillator is scanned using delay stage 2. BS: 50/50 beamsplitter.}
    \label{fig: exp_scheme}
\end{figure}

Figure~\ref{fig: exp_scheme} illustrates the experimental setup for the generation of ultrafast squeezed light with attosecond control and frequency-resolved balanced homodyne detection. Squeezed light is generated by a Degenerate Four-Wave Mixing (DFWM) interaction of three non-collinear near-infrared (NIR) femtosecond laser pulses with a $100~\mu m$ thick magnesium oxide sample. The three input pulses consist of a phase-locked gate pair and a probe arranged in a folded BOXCARS geometry. The signal is emitted along the phase-matched direction $\bm{k}_s = \bm{k}_1-\bm{k}_2+\bm{k}_3$ and is spatially separated from all input beams. The signal mode is not seeded. It builds up from the vacuum through the third-order polarization, which requires all three input fields. Without the probe, there is no signal in the phase-matched direction. The probe therefore plays two roles. It is one of the three fields that drive the wave mixing and its sub-cycle delay relative to the phase-locked gates sets the total instantaneous field in the focal volume, which modulates the nonlinear response \cite{walz2025}. The generated signal is sent to a balanced homodyne detection setup that incorporates a diffraction grating and a standard CMOS camera for frequency-resolved detection (see Methods). To perform frequency-resolved balanced homodyne detection, the single-shot spectra from the two arms of the homodyne detection setup are subtracted for each wavelength. Although the pixels on the camera corresponding to the same wavelength in each spectrum are similar, they do not have well-matched noise characteristics. As a result, balanced detection sufficient to reach the quantum noise limit of the laser cannot be achieved directly. We overcome this limitation by utilizing a strong-field-driven modulation of the material's nonlinear response~\cite{walz2025} and filtering the squeezed light signal based on this modulation, as described below.

\subsection*{Nonlinear Response Modulation}
In perturbative nonlinear optics, the nonlinear response functions of a material (nonlinear susceptibilities) depend only on material properties and are independent of external fields. However, when moderately strong laser fields similar to field strengths used in solid-state high-order harmonic generation (HHG) \cite{Ghimire2011}, which are typically in the $10^{11} - 10^{12}$ W/cm$^2$ range, are applied, the nonlinear susceptibilities are modified by the external fields. When generating squeezed light from DFWM, the properties of squeezed light such as the degree of squeezing are directly related to third-order susceptibility \cite{Orszag2016, Agarwal_2012}. Hence, a strong-field-driven modulation of this susceptibility leads to attosecond control over the generated squeezed light. We note that periodic modulation of pump light in a nonlinear optical cavity has been proposed to generate highly squeezed light \cite{adamyan2015} but field driven nonlinear response modulation has not been previously explored. In our work, we show that in addition to attosecond control, quantum noise level can be reached in frequency-resolved balanced homodyne detection measurements using a CMOS camera by leveraging this nonlinear response modulation.

When the time delay between the phase-locked gate pulses and the probe pulse is varied on sub-cycle time scales, the total electric field at a given point in the focal volume modulates because of constructive and destructive interference. This modulation of the total field modulates the nonlinear response of the material \cite{walz2025}, resulting in modulation of the generated squeezed light. Figure~\ref{fig: exp_scheme} inset (bottom right corner) shows the homodyne signal as a function of time delay ($\tau$) for a fixed relative phase ($\phi_{LO}$) of the local oscillator with respect to the signal. We use a low-pass filter that filters out any time-delay frequencies higher than $1\omega_0$, where $\omega_0$ is the frequency corresponding to the central wavelength of 808 nm. The details of this filtering process are explained in the Methods section and in the supplementary material.

\subsection*{Quantum Tomography and Attosecond Control of Quadrature Squeezing}

\begin{figure}[h]
    \centering
    \includegraphics[width=\linewidth]{Final_figure/Figure3_v10.pdf}
    \caption{Attosecond control of squeezing and calibration of the quantum noise. Normalized Wigner distribution functions obtained from the homodyne signal at time delays (a) $\tau = 5.4$ fs and (b) $\tau = 10.2$ fs. The corresponding NRM-filtered Wigner functions are shown in (c) and (d). The NRM-filtered Wigner functions at 808 nm are fit to a 2D Gaussian at each time delay. (e) The extracted quadrature variances as a function of time delay. Red and black arrows in (c) and (d) indicate the corresponding variances in (e). $X_{\theta}$ and $P_{\theta}$ are generalized quadratures. A blue dashed line separates the rapidly varying region (Region 1) from the constant shot noise level (SNL) region (Region 2) (see main text). (f) Oscillation of the maximum and minimum variances with time delay at different wavelengths within the pulse bandwidth. The oscillations at different wavelengths are synchronized in phase. (g) Relative intensity noise (RIN) obtained from the power spectral density of the camera homodyne signal for the unfiltered coherent state (magenta), the NRM-filtered coherent state (green), and the NRM-filtered squeezed state (blue) at $\tau = 7.6$~fs (coherent) and $\tau = 10.2$~fs (squeezed). (h) Quadrature noise of the coherent state from the homodyne measurement compared with the laser shot noise measured with standard balanced photodiode detectors. Their overlap shows that the coherent state RIN in (g) is at the shot-noise level and that the measured squeezing is below the shot-noise limit. (i) Posterior distribution of the squeezing at 808 nm and $\tau = 10.2$~fs from a Bayesian inference over Gaussian states (see supplementary material). The mean is $-1.62$~dB and the shaded region marks the $90\%$ credible interval of $[-1.67, -1.58]$~dB.}
    \label{fig: WDF}
\end{figure}
%\textcolor{blue}{(is this the $Q_{\theta}$ variance?)}
%\textcolor{blue}{(Consider making the color bar uniform across panels (a)-(d), and only showing one colorbar.)}

The NRM-filtered homodyne detection technique is employed to directly measure the $\phi_{LO}$ phase-dependent generalized quadratures of the DFWM output signal. Using optical homodyne tomography, we reconstruct the Wigner distribution function of the DFWM signal \cite{alvarez2023, Wu98}. We use the inverse Radon transformation to obtain the Wigner functions from the measurement of homodyne signal. Details of the reconstruction algorithm are provided in the Methods section. The Wigner distributions were constructed for all time delays ($\tau$) using both raw (unfiltered) and NRM-filtered data. Figures~\ref{fig: WDF} (a) and (b) show the Wigner functions corresponding to time delays of $\tau = 5.4~\mathrm{fs}$ and $10.2~\mathrm{fs}$ respectively for 808 nm, obtained from the data without NRM filtering, while Figs.~\ref{fig: WDF} (c) and (d) show the corresponding NRM-filtered results. The Wigner functions are shown for the central wavelength (808 nm) of the output signal. Amplitude squeezing and phase squeezing are prominently observed. The reconstructed Wigner functions also show weak negative regions. Negativity of the Wigner function is a standard indicator of non-classical light \cite{Kenfack2004}. However, the inverse Radon transform can produce weak negative features as reconstruction artifacts. We therefore do not use this negativity as evidence of non-classicality. The non-classical nature of the light is instead established by the calibrated sub-shot-noise quadrature variances presented below. The Gaussian character of the state is tested directly on the measured quadrature data later in the text. The shift in the centroid of the Wigner distribution with time delay arises from the change in the global phase of the DFWM signal with respect to the local oscillator, which cannot be compensated for in the current experimental configuration. We emphasize that this global phase does not affect our conclusions. A delay of the local oscillator, or a delay-induced global phase of the signal, acts on the quantum state as a rigid rotation in phase space. Under such a rotation, the noise ellipse and the displacement vector rotate together. Two quantities are invariant under this rotation. The first is the pair of eigenvalues of the covariance matrix, which set the degree of squeezing. The second is the orientation of the noise ellipse relative to the mean field, which distinguishes amplitude squeezing from phase squeezing. Our claim of attosecond control rests only on these invariant quantities. Figures \ref{fig: WDF} (e) and (f) show that the minimum and maximum quadrature variances themselves modulate with time delay at the optical period. Figures \ref{fig: WDF} (c) and (d) show the state switching between phase squeezed and amplitude squeezed relative to its own mean field. No phase shift of the local oscillator or of the signal field can change these quantities. The time delay is therefore not simply shifting a phase. It modulates the generation process itself through the field-driven nonlinear response $\chi^{(3)}(E)$ \cite{walz2025}. To quantify the degree of squeezing, we fit a two-dimensional Gaussian function to the Wigner functions and extract the maximum and minimum variances at each time delay for generalized quadratures $X_{\theta}$ and $P_{\theta}$. Figure~\ref{fig: WDF} (e) shows the modulation of the degree of squeezing as a function of time delay extracted using the fitting procedure. Figure~\ref{fig: WDF} (e) is divided into two regions, Region 1 and Region 2. We do not observe a constant shot noise level (point of intersection of the red and black curves) in Region 1 likely due to experimental instability as well as rapid variation in signal intensity near zero delay as compared to Region 2. In Region 2 the shot noise is measured to be constant and well defined. Using homodyne detection of the nonlinear DFWM process driven by ultrafast femtosecond pulses, we have demonstrated that the squeezing of the output signal can be controlled with unprecedented temporal resolution on the sub-cycle attosecond scale via the controlled modulation of third-order nonlinear susceptibility, $\chi^{(3)}$, of the material. Such fine control over the properties of squeezed light could have major applications in attosecond quantum optics where quantum light is being used to drive processes such as HHG \cite{tzur2025_fluctuation, Tzur2025} and multiphoton ionization \cite{Dahan2021}.

%because the shot-noise level is not the same in these regions. As the time delay is varied in the experiment, the intensity  of the generated squeezed light changes and this variation is rapid closer to zero delay. As a result, the shot noise level (SNL), which depends on the intensity of the signal also varies rapidly in Region 1. However, in Region 2, the shot-noise level is approximately constant and corresponds to the point of intersection of the variance curves in Fig.~\ref{fig: WDF} (e), which represents a coherent state where the quadrature variances are equal.

To provide further evidence that we are observing squeezing below the quantum noise limit, we performed a quantum noise analysis of the measured homodyne signal using the measured power spectral density (PSD). The relative intensity noise (RIN) was calculated from the raw, unfiltered data for a near-coherent state at $\tau = 7.6~\mathrm{fs}$, where the variances in both quadratures are nearly equal, and is plotted in Fig.~\ref{fig: WDF} (g) (magenta). The corresponding RIN for the NRM-filtered data of the near-coherent state is shown in green in Fig.~\ref{fig: WDF} (g). We show that by applying the NRM filter, the classical noise in our measurements is reduced by a few decibels (dB). The RIN of the signal corresponding to maximum squeezing (at $\tau = 10.2~\mathrm{fs}$) is plotted in blue in Fig.~\ref{fig: WDF} (g). This shows that the quantum noise of the squeezed state lies several dB below that of the coherent state. Since direct shot-noise measurements are not possible using a CMOS camera, we repeat the experiment using a balanced photodiode detector to measure the homodyne signal. The quadrature noise was measured over the full spectral range of the DFWM signal. The measured quadrature noise for the coherent state at $\tau = 7.6~\mathrm{fs}$ overlaps with the laser’s shot-noise limit, as shown in Fig.~\ref{fig: WDF} (h), confirming that the noise in the near-coherent state approaches the fundamental shot-noise limit of the laser pulse. Since Fig.~\ref{fig: WDF} (g) clearly shows that the squeezed-state noise lies below that of the near-coherent state, the squeezed-state noise lies below this shot-noise-referenced floor. To estimate the maximum degree of squeezing achieved in this experiment, we integrate the RIN of the NRM-filtered coherent and squeezed states around the $10~\mathrm{Hz}$ frequency region (see the dashed rectangular box in Fig.~\ref{fig: WDF} (g)). In this low electronic-noise frequency region, the RIN of the squeezed light lies $\sim 5$ dB below that of the identically filtered near-coherent state. This band-limited comparison shows the noise reduction relative to the reference. It is not a calibrated measure of the degree of squeezing. The RIN analysis establishes squeezing below the shot-noise level but does not by itself provide an uncertainty on the squeezing value. To obtain a calibrated value with uncertainty, we perform a Bayesian inference over the family of Gaussian states (see supplementary material). The covariance matrix is parametrized in Williamson form. Every state visited by the sampler therefore satisfies the uncertainty principle and is a valid quantum state. The inference uses only the NRM-filtered single-shot, mean-subtracted quadrature values and involves no inverse Radon transform and no Wigner function. The resulting posterior distribution of the squeezing at 808 nm and $\tau = 10.2~\mathrm{fs}$ is shown in Fig.~\ref{fig: WDF} (i). The posterior mean is $-1.62$ dB with a $90\%$ credible interval of $[-1.67,-1.58]$ dB. All posterior samples correspond to squeezing below the shot-noise level.

The measured sub-shot-noise squeezing is modest compared to state-of-the-art cavity-based continuous-wave sources, which reach 15 dB \cite{Vahlbruch2016}. The main contribution of this work is a new control mechanism and a multimode measurement method that operate at high peak power, not at a record-squeezing level. Several factors limit the measured value. The reported squeezing is not corrected for detection efficiency and thus represents a lower bound on the source's squeezing (see Methods). The classical noise residue retained after NRM filtering biases the measurement toward less squeezing. The achievable nonlinearity is set by the 100 $\mu m$ interaction length in MgO and by input intensities bounded by the damage threshold. The depth of the nonlinear response modulation sets the range of control. Realistic routes to higher squeezing include detectors with higher quantum efficiency and lower read noise, longer interaction lengths such as in gas-filled hollow-core fibers or in media with larger $\chi^{(3)}$, and application of the same modulation approach to $\chi^{(2)}$ media such as periodically poled lithium niobate.

\subsection*{Multi-mode Quadrature Correlations}
As discussed earlier, one of the key advantages of using a spectrometer for detection is that it enables frequency-resolved measurement of the homodyne signal. This allows us to access the quantum state of the emitted signal across different frequency modes. Moreover, because all frequency components are recorded simultaneously within a single shot, it is possible to extract quadrature correlations between these frequency modes. We thus analyze the homodyne signal across different frequency components and perform optical homodyne tomography to reconstruct the corresponding Wigner functions. The Gaussian fitting procedure is then applied to extract the quadrature variances for each frequency mode. Figure~\ref{fig: WDF} (f) shows the modulation of the quadrature variance as a function of time delay for several selected wavelengths near the central wavelength $\lambda = 808~\mathrm{nm}$. As we deviate from the central wavelength, the degree of squeezing reduces, however, the variance oscillations are completely in-phase, demonstrating strong correlations between different frequency modes.

%To the best of our knowledge, this represents the first experimental realization of a broadband, frequency-resolved quantum state measurement.

\begin{figure}[h]
    \centering
    \includegraphics[width=0.9\linewidth]{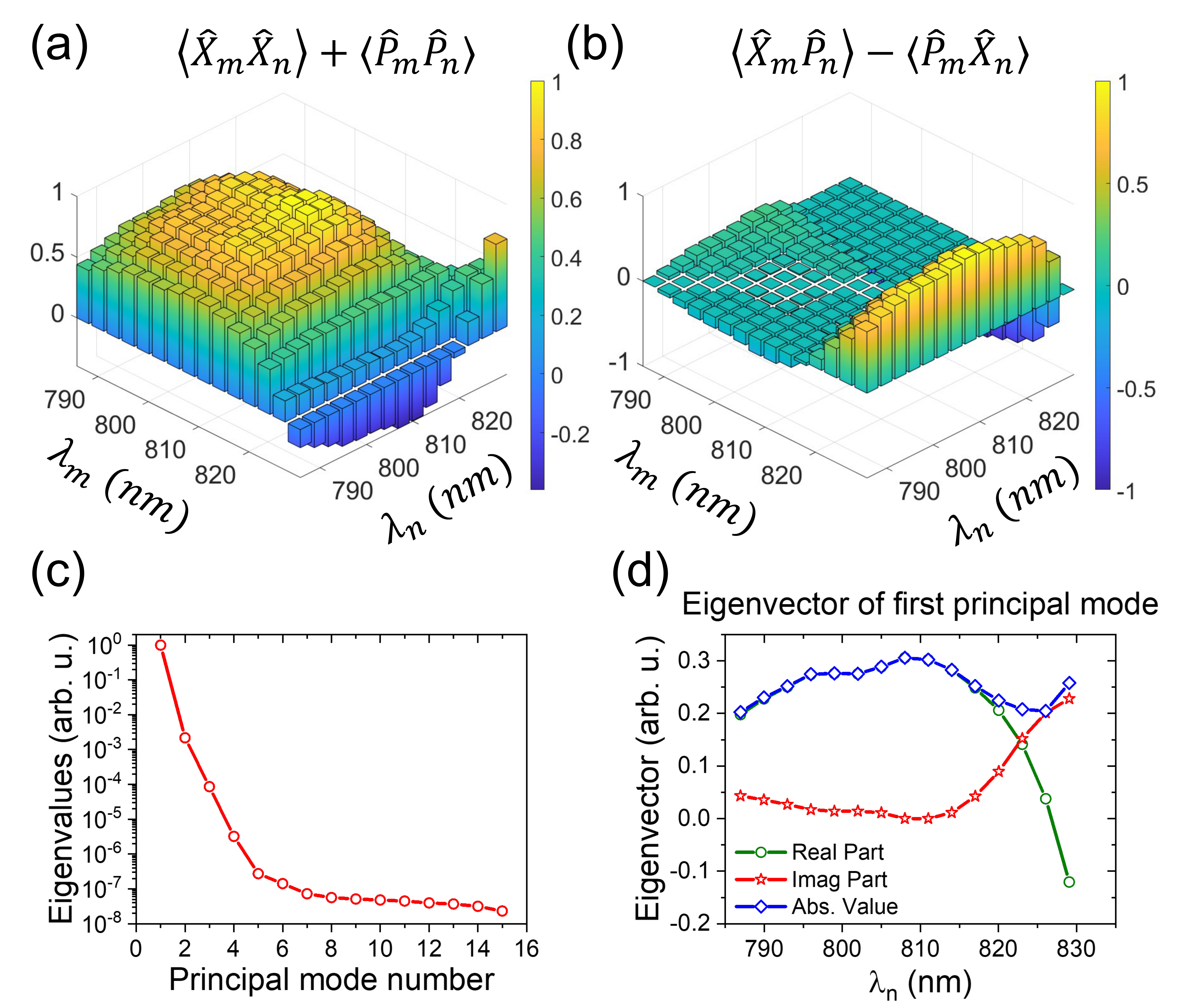}
    \caption{The correlation functions (a) $\langle X_m X_n \rangle$ + $\langle P_m P_n\rangle$ and (b)  $\langle X_m P_n \rangle$ - $\langle P_m X_n\rangle$ are shown, where $m$ and $n$ denote the frequency modes. Here, $X$ represents the quadrature along the direction of maximum squeezing, and $P$ is the conjugate quadrature orthogonal to $X$, evaluated for the Wigner function at $\tau = 10.2~\mathrm{fs}$. From these correlation functions, the coherency matrix is constructed (see Eq.~\ref{coherency_matrix}). The eigenvalues of the coherency matrix, obtained via diagonalization, are shown in (c). The eigenvector corresponding to the dominant first principal mode, presented in (d), reveals that this mode constitutes a linear combination of multiple frequency components with approximately equal weights.}
    \label{fig: correlations}
\end{figure}

For a multimode quantum state, the information on quantum correlations is contained within the quantum coherency and quantum covariance matrices. The coherency matrix is defined as \cite{Fabre2020}
\begin{equation} \label{eq:coherency_matrix}
    (\Gamma^{(1)})_{mn} = \langle \hat{a}^{\dagger}_m \hat{a}_n\rangle\, 
\end{equation}
where $\hat{a}^{\dagger}_m (\hat{a}_m)$ is the creation (annihilation) operator of a photon in frequency mode $m$. The second-order moments of the field quadratures are contained in the covariance matrix, defined  \cite{Fabre2020} 
\begin{equation}
    \Sigma = 
    \begin{pmatrix}
        \langle \hat{X}_m\hat{X}_n\rangle & \langle \hat{X}_m\hat{P}_n\rangle \\
        \langle \hat{P}_m\hat{X}_n\rangle & \langle \hat{P}_m\hat{P}_n\rangle
    \end{pmatrix}\, 
\end{equation}
In our homodyne experiment, we directly measure the quadratures $X_m$ and $P_m$ of different frequency modes simultaneously. Consequently, the complete covariance matrix can be reconstructed from the experimentally acquired data. The coherency matrix can further be calculated from the quadrature correlations using the relation~\cite{Fabre2020}
\begin{equation}\label{coherency_matrix}
    (\Gamma^{(1)})_{mn} = \frac{1}{4}\left[\langle \hat{X}_m\hat{X}_n\rangle + \langle \hat{P}_m\hat{P}_n\rangle + i (\langle \hat{X}_m\hat{P}_n\rangle - \langle \hat{P}_m\hat{X}_n\rangle)\right]\, 
\end{equation}

%The real part of the coherency matrix is shown in Fig.~\ref{fig: correlations} (c), which is the sum of Fig.~\ref{fig: correlations}(a) and (b), while its imaginary part, shown in Fig.~\ref{fig: correlations}(d), captures the difference between the off-diagonal covariance matrix elements. 

%The trace of the un-normalized coherency matrix provides the mean total photon number of the quantum state, estimated to be on the order of $\sim 10^{10}$ photons, which is consistent with typical photon numbers observed in a single pulse of the DFWM signals generated by femtosecond lasers.

Fig.~\ref{fig: correlations} (a) and (b) show the real and imaginary part of the coherency matrix. Diagonalization of the coherency matrix yields the principal eigenmodes. The eigenvalues of the coherency matrix are shown in Fig.~\ref{fig: correlations} (c) and represent the mean photon numbers in each principal mode. Before the interaction with the material, the coherent state is confined to a single principal mode. Following the DFWM interaction, mode mixing occurs, leading to the generation of a multimode squeezed state of light. The large eigenvalue of the first principal mode seen in Figure~\ref{fig: correlations} (c) indicates that although intermodal coupling is present, the output quantum field remains predominantly governed by the first principal mode. The eigenvector corresponding to the first principal mode shown in Fig.~\ref{fig: correlations} (d) indicates that the first principal mode is a linear combination of all frequencies with approximately equal weights.

\begin{figure}[h]
    \centering
    \includegraphics[width=\linewidth]{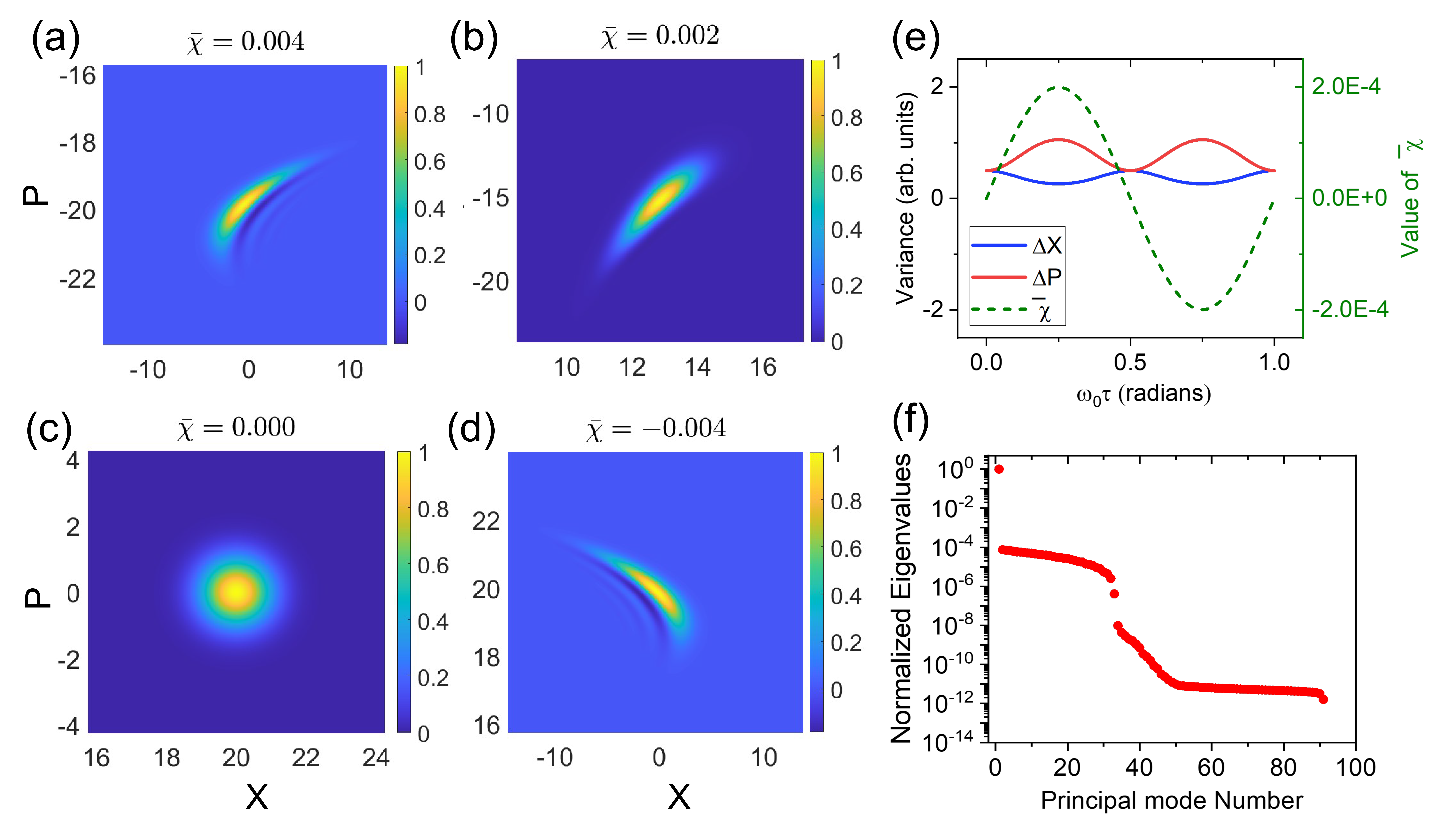}
    \caption{The Wigner distributions (see Eq.~\ref{theory_WDF}) calculated using the nonlinear Schrödinger equation (see Eq.~\ref{NLSE}) are shown for different values of the dimensionless coupling constant $\bar{\chi}$ (a) $0.004$, (b) $0.002$, (c) $0$, and (d) $-0.004$. The values of $\bar\chi$ here are larger than those realized in the experiment and are chosen to make the dependence on the coupling strength more visible. Panel (e) illustrates that modulation of $\bar{\chi}$ leads to corresponding modulation in the variances of specific $X$ and $P$ quadratures. The values of $\bar{\chi}$ in (e) are lower than those used in (a) - (d). These lower values are used to show a simple modulation of the variances as a function of $\bar{\chi}$. At higher values of $\bar{\chi}$, the modulation develops higher frequencies (see supplementary material). The eigenvalues of the calculated coherency matrix are shown in panel (f), indicating that the photon population is predominantly confined to the first principal mode, with only minor contributions from higher-order modes. This result is consistent with the principal mode decomposition obtained from the experimentally measured data.}
    \label{fig:theory_plots}
\end{figure}

It has been well established that the quantum properties of light are influenced by the nonlinear response of the material through which it propagates. In particular, for nonlinear optical processes such as degenerate four-wave mixing, non-classical features including the squeezing parameter are directly determined by the material’s third-order susceptibility \cite{Agarwal_2012, Drummond2014, Orszag2016}. However, when strong femtosecond laser fields participate in the wave-mixing process, the electronic configuration of the medium is non-perturbatively modified in the presence of the field. In solid-state systems, this effect arises from strong-field–induced modulation of the electronic band structure \cite{gruzdev2018, uzan2022, lucchini2016}. In a recent study, it has been shown that the third-order susceptibility is modulated by changing the time delay between the input femtosecond pulses \cite{walz2025}. 

Though a complete simulation of the quantum optical interaction in the presence of strong fields is currently very challenging, we phenomenologically introduce the modulation of susceptibility into an existing quantum optical model to describe the quantum state of the output signal. We simulate the quantum state of light of the output signal in the four-wave mixing process by solving the effective one-dimensional nonlinear Schr\"odinger's equation for the photon field propagating through a nonlinear medium (see Methods). We then parameterize the dimensionless coupling constant $\bar{\chi}$ (proportional to third-order susceptibility) with time delay between the input pulses.
% We have performed numerical simulations of the quantum state of light in a four-wave mixing process to support our experimental observation that the degree of squeezing is modulated by the field-driven variation of the nonlinear susceptibility. Specifically, we solved the nonlinear Schrödinger equation (see Eq.~\ref{NLSE} in Methods), where the nonlinear coupling constant $\chi_e$ is proportional to the third-order nonlinear susceptibility $\chi^{(3)}$.
The full numerical solution of the quantum light field after propagation through the crystal is well approximated by the nonlinear evolution of its first principal mode, which can be solved analytically (see Methods).
The computed Wigner distribution functions display negative components of the Wigner function characteristic of non-classical light (Fig.~\ref{fig:theory_plots} (a) - (d)). The dependence on $\bar{\chi}$ clearly illustrates that the degree and type of squeezing varies with the coupling strength. The values of $\bar{\chi}$ used in Fig.~\ref{fig:theory_plots} (a)-(d) are larger than those realized in the experiment. These values are chosen to make the state's dependence on the coupling strength clearly visible. At the coupling strength that reproduces the measured variances, the model predicts a state that is Gaussian to a very good approximation. Furthermore, we show in Fig.~\ref{fig:theory_plots} (e) that a periodic modulation of $\bar{\chi}$ modulates the variance of quadratures $X$ and $P$. In our calculations, $\bar{\chi}$ is independent of frequency because the third-order susceptibility in the experiment is expected to be constant over the bandwidth of our laser pulse for a large band-gap dielectric such as MgO. This assumption is also consistent with our experimentally extracted frequency-frequency correlations that are nearly constant over the full bandwidth (see Fig. \ref{fig: correlations}). Our effective model thus captures the main emergent quantum light features, including the single-mode character and quadrature variance oscillations, allowing a simple, semi-analytic analysis of the results. A complete quantitative model would require a quantum optical description of strong field driven processes in a dielectric.

\begin{figure}[h]
    \centering
    \includegraphics[width=\linewidth]{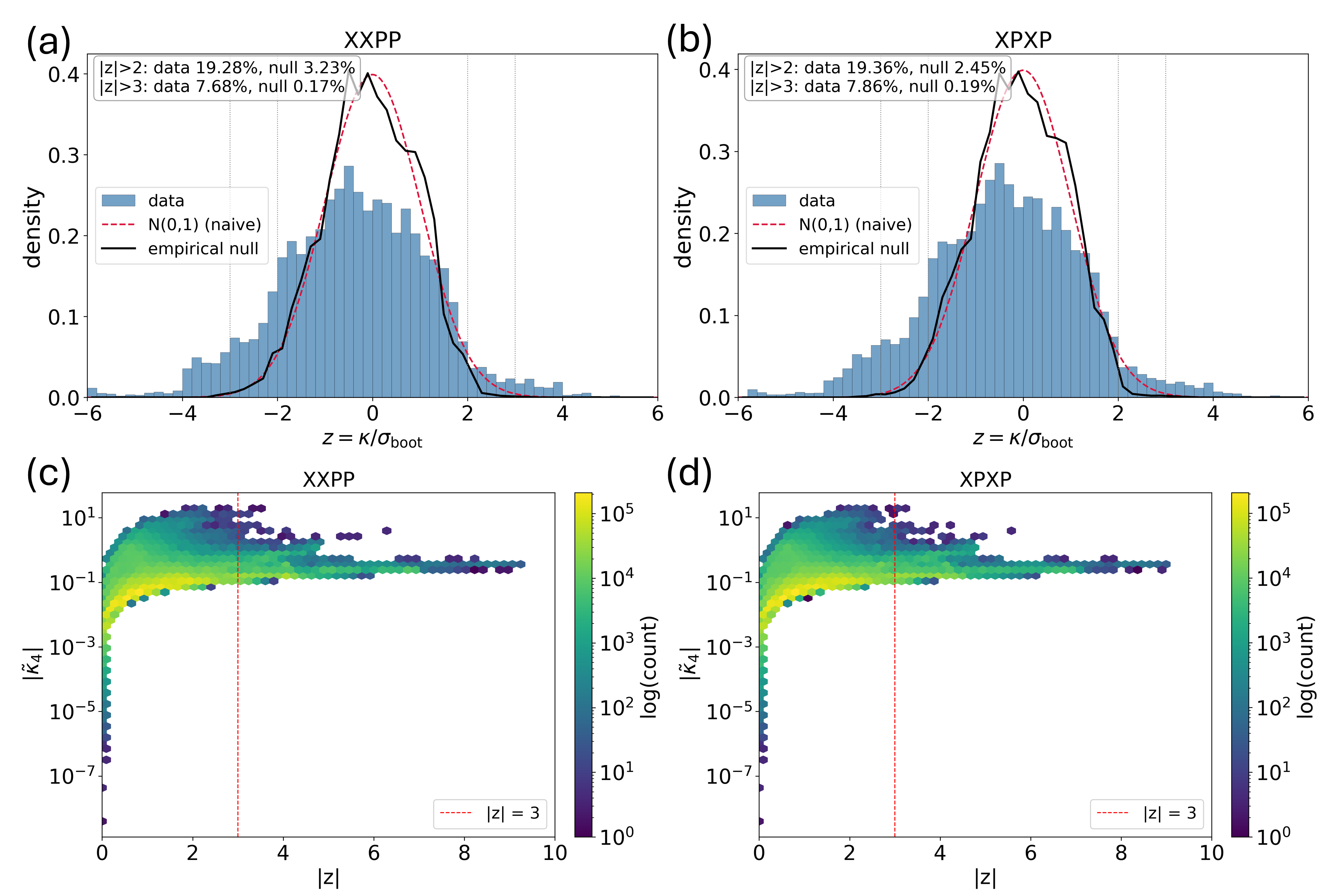}
    \caption{The histogram of the bootstrapped z-score is presented for (a) $\kappa_4(\delta X,\delta X, \delta P, \delta P)$ and (b) $\kappa_4(\delta X,\delta P, \delta X, \delta P)$. The red dashed curve represents the normal distribution $\mathcal{N}(0, 1)$, while the black solid curve represents the empirical null, simulated using a surrogate Gaussian state (see the main text and supplementary material for details). A substantial proportion of entries outside the $3\sigma$ region ($|z|>3$) indicates that the measured quantum state exhibits non-Gaussian characteristics. The joint density plots of the z-score versus the standardized cumulant ($\tilde \kappa_4$) are shown in (c) and (d). These plots demonstrate that entries beyond the $3\sigma$ region cluster around $|\tilde \kappa_4| \approx 0.1$, suggesting that the measured state is approximately 90\% Gaussian and 10\% non-Gaussian.}
    \label{fig:cumulant}
\end{figure}

The second-order moment shown in Figure~\ref{fig: correlations} quantifies the Gaussian character of the measured quantum state. However, the introduction of nonlinear interactions within the system can generate non-Gaussian states \cite{Mlmer2006, Tang2015, Lezama2023, Zhang2025, Rasputnyi2025}. When an intense laser interacts with a nonlinear medium, the squeezing process becomes distorted, transforming the Gaussian distribution of the quantum state into an exotic, non-Gaussian form \cite{Rasputnyi2025}. To detect non-Gaussianity in the measured squeezed state, the fourth-order cumulant of the quadrature noise $\delta X$ and $\delta P$ is evaluated. The general expression for the fourth-order cumulant is given by,
\begin{align}\label{eq:kappa_4}
    \kappa_4 (A_{\alpha},B_{\beta},C_{\gamma},D_{\delta})
    &= \langle A_{\alpha}B_{\beta}C_{\gamma}D_{\delta}\rangle
       - \langle A_{\alpha}B_{\beta}\rangle \langle C_{\gamma}D_{\delta} \rangle \nonumber\\
    &\quad - \langle A_{\alpha}C_{\gamma}\rangle \langle B_{\beta}D_{\delta} \rangle
       - \langle A_{\alpha}D_{\delta}\rangle \langle B_{\beta}C_{\gamma} \rangle,
\end{align}
In this context, $A, B, C, D$ each denote either the $\delta X$ or $\delta P$ quadrature, and the indices $\alpha, \beta, \gamma, \delta$ correspond to different frequency modes within the broadband pulse. Equation~\ref{eq:kappa_4} demonstrates that the cumulant quantifies the deviation of the fourth-order moment from the second-order moment. The bootstrapping technique is employed to estimate $\kappa_4$ while accounting for statistical noise in the data (see the supplementary material for details of the analysis). The z-score is then evaluated using the bootstrapped standard deviation $\sigma_{\text{boot}}$ as follows:
\begin{equation}\label{eq:z-score}
    z = \frac{\kappa_4}{\sigma_{\text{boot}}}.
\end{equation}
Figure~\ref{fig:cumulant} (a) and (b) present the histogram of the z-score evaluated using Equation~\ref{eq:z-score} for $\kappa_4(\delta X,\delta X, \delta P, \delta P)$ and $\kappa_4(\delta X,\delta P, \delta X, \delta P)$, respectively. For a perfectly Gaussian measured state, $\kappa_4=0$ within statistical noise, and the bootstrap histogram would follow a normal distribution $\mathcal{N}(0,1)$. However, the observed histogram deviates significantly from a normal distribution, with more than 7\% of the entries falling beyond $|z|=3\sigma$ region, compared to only 0.27\% for a normal distribution. An empirical null distribution was also simulated using a surrogate Gaussian state. The empirical null closely follows a normal distribution, indicating that the $\kappa_4$ plug-in estimator exhibits negligible finite-sample bias for repeated indices (see supplementary material for details). While the z-score serves as a useful indicator of non-Gaussianity, it does not provide a quantitative measure of its extent in the quantum state. Therefore, the standardized fourth-order cumulant is also evaluated, defined as follows:
\begin{equation}
    \tilde{\kappa}_4(\alpha,\beta,\gamma,\delta) =
    \frac{\kappa_4(\alpha,\beta,\gamma,\delta)}
         {\langle A_{\alpha}B_{\beta}\rangle \langle C_{\gamma}D_{\delta}\rangle
        + \langle A_{\alpha}C_{\gamma}\rangle \langle B_{\beta}D_{\delta}\rangle
        + \langle A_{\alpha}D_{\delta}\rangle \langle B_{\beta}C_{\gamma}\rangle}.
\end{equation}
Figure~\ref{fig:cumulant} (c) and (d) display the joint density plot between the z-score and $\tilde\kappa_4$. Entries lying outside the $3\sigma$ region ($|z|>3$) predominantly cluster near $|\tilde \kappa_4| \approx 0.1$. This result indicates that the most strongly squeezed state exhibits approximately 10\% non-Gaussianity. This result connects the measurement to the effective model in a specific way. At the experimentally relevant coupling strength, the model predicts a Gaussian state. The measured state is indeed Gaussian to leading order, in agreement with the model. The ten percent non-Gaussian correction is a real feature that the model does not produce. A plausible origin is the non-perturbative nature of the strong-field interaction. The model varies only the strength of a perturbative cubic response. In the experiment, the nonlinear response is an instantaneous function of the total field, $\chi^{(3)}(E)$, due to field-driven modulation of the band structure \cite{walz2025}. This effect is not equivalent to adding higher-order susceptibilities (see supplementary material). The measured weak non-Gaussianity therefore identifies a concrete target for future quantum optical models of strong-field driven dielectrics. At the variance level, a ten percent effect corresponds to approximately 0.4 dB. The sub-shot-noise conclusion is unchanged by this correction.

In conclusion, we have experimentally demonstrated sub-cycle, attosecond control of ultrafast femtosecond squeezed light generation from a nonlinear optical process. We have measured quadrature correlations between different frequency modes simultaneously across the full bandwidth of the femtosecond pulse. This was made possible by a fundamentally new approach towards generation and measurement of field quadrature squeezing by leveraging strong-field driven modulation of the nonlinear response in a large band-gap dielectric \cite{walz2025}, which allows us to reach the quantum noise limit using an unbalanced CMOS detector. From our experimental measurements, we directly construct Wigner distribution functions representing the full quantum state of the DFWM signal, which prominently shows control over the degree and type of squeezing as the time delay is varied with attosecond resolution. Furthermore, using the simultaneous measurement of field quadratures in multiple frequency modes, we have constructed the coherency matrix and determined that a single principal mode is dominant in the generated squeezed light. The detailed theoretical calculations based on solving the nonlinear Schr\"odinger equation including a phenomenological variation of the nonlinear response show qualitative agreement with experimental results. Our results connect to three directions. First, a growing body of work drives strong-field processes such as high-harmonic generation and multiphoton ionization with quantum light, where the outcome depends on the photon statistics of the driver \cite{Rasputnyi2024,Tzur2025,tzur2025_fluctuation, Dahan2021, EvenTzur2023,long2025}. A source whose degree and type of squeezing can be switched on sub-cycle time scales provides a control knob for such experiments. The quantum statistics of the driving field become a tunable parameter synchronized to the sub-cycle dynamics being driven. Second, frequency-multimode squeezed light is the resource underlying continuous-variable quantum information protocols \cite{RevModPhys.77.513,Fabre2020,Roslund2013}. Our measurement records all frequency modes of each pulse simultaneously and extracts the full coherency matrix. Combined with fast control of the generation process, this is a step toward multimode squeezed sources whose mode structure and squeezing can be reconfigured dynamically rather than fixed by a cavity. Third, because the squeezing parameters are set by the instantaneous nonlinear response, attosecond-scale dynamics of that response can imprint themselves on the quantum correlations of the emitted field. Multimode quadrature correlation measurements at the quantum noise limit, as demonstrated here, provide the transduction required to study such dynamics.

\section*{Methods}\label{sec:Methods}

\subsection*{Experimental Setup}
The generation and measurement scheme of our squeezed light signal is shown in Fig.~\ref{fig: exp_scheme} and employs degenerate four-wave mixing (DFWM) coupled with a strong local oscillator to achieve homodyne detection. 50 fs pulses with a center wavelength 808 nm from a 1 kHz repetition rate Ti:Sapph laser are first split using a 50/50 beam splitter. The reflected arm becomes the local oscillator and is sent to delay stage 2 to control relative phase, $\phi_{LO}$, between the signal (squeezed light) and the local oscillator in 100 steps per optical cycle with a step size of 27 attoseconds. The transmitted arm is further separated into two phase-locked gate beams and a probe beam using a beam splitter and a two-hole mask. Time delay, $\tau$, between the two gate beams and the probe beam is controlled via delay stage 1 with a delay resolution of approximately 200 attoseconds. The probe and gate beams, with average intensities of $\sim 2~TW/cm^{2}$ and $\sim 1~TW/cm^{2}$, respectively, interact with a 100 $\mu$m thick sample of MgO (100) via a BOXCARS\cite{Shirley1980} configuration to generate the squeezed signal. All input beams are spatially filtered using an iris. The resulting signal beam is spatially isolated from the input beams and coupled to the local oscillator via a typical homodyne measurement beam path with a power ratio of approximately 40:1. The DFWM signal is bright, with an estimated mean photon number of approximately $10^8$ per pulse. The coherent displacement is therefore the dominant feature of the quantum state. As seen in Fig.~\ref{fig: exp_scheme}, we use a diffraction grating with 1500 grooves/mm and a CMOS camera to make a spectrally-resolved homodyne measurement. For this two-dimensional scan ($\tau$ and $\phi_{LO}$), 1000 single shots of both paths are captured simultaneously at each delay position. The camera is triggered at 100 Hz, and each frame records a single laser pulse, corresponding to every tenth pulse of the 1 kHz train. The exposure time is short compared to the 1 ms pulse spacing. No averaging over consecutive pulses occurs, and the quadrature statistics are single-pulse statistics. Our spectrometer has a calibrated spectral resolution of 0.06 nm per pixel. Isolation of individual frequency modes, as well as implementation of nonlinear response modulation filtering, allows this measurement with a CMOS detector to reach the shot noise limit without the necessity of balanced integrated circuitry normally seen in a balanced photodiode system of homodyne detection\cite{Bachor2019}. The overall detection efficiency of the homodyne measurement is $\eta_{tot} = \eta_{cam} V^2 T \approx 0.29$. Here $ \eta_{cam} \approx 0.38$ is the quantum efficiency of the camera at 808 nm and $T \approx 0.95$ is the optical transmission from the sample to the detector. The visibility V between the local oscillator and the signal at 808 nm is extracted from the mean homodyne fringe at the near-coherent reference delay, where the signal is brightest and the extraction most robust, giving $V = 0.89$ (see supplementary material). At the squeezed-state delay the extracted fringe amplitude is consistent with unit visibility within the intensity-normalization uncertainty. Since the mode overlap is set by the beam geometry and does not depend on the sub-cycle delay, the reference-delay value is used for the visibility. The measured squeezing of $-1.62$ dB requires $\eta_{tot} \geq 0.31$ for any physical input state, within approximately 0.2 dB of the measured $\eta_{tot}$. This small difference is within the systematic uncertainty of the visibility determination (see supplementary material). The squeezing values reported in this work are not corrected for detection efficiency. They are directly measured values and represent lower bounds on the squeezing of the source.

\subsection*{Optical Homodyne Tomography}
We sampled 1000 laser shots of the homodyne signal for each $\phi_{LO}$ and $\tau$ and constructed the probability distribution function (PDF) of the homodyne signal ($i_-$). To reconstruct the Wigner function, an inverse Radon transformation is performed \cite{alvarez2023, Wu98},
\begin{equation}
\label{Radon_trans}
    W_{\tau}(Q, P) = \frac{1}{4\pi^2}\iint \int_0^\pi pr_{\tau}(s, \phi) \exp{\left[-i\eta\left(s-Q\cos{\phi} - P\sin{\phi}\right)\right] d\eta ds d\phi}\, 
\end{equation}
where $pr_{\tau}(s, \phi)$ is the PDF of the homodyne signal in various phases of $\phi = \phi_{LO}$, and $\tau$ is the time delay between gate and probe pulses, which parameterizes the PDF. $s$ is the measured quadrature value and $\eta$ is the variable which defines the kernel of the Radon transformation. Figures~\ref{fig: WDF} (a)-(d) show the Wigner distribution functions of the DFWM signal in generalized quadrature $X$ and $P$ for two different values of time delay $\tau$ using raw and NRM-filtered data (see supplementary material for the NRM filtering procedure). The Wigner functions in Fig.~\ref{fig: WDF} are plotted in a window centered on the distribution. The large coherent displacement of the state is therefore not visible on the plotted axes. To estimate the amplitude and phase variances of the constructed Wigner function, we fit a generalized 2D Gaussian function \cite{navarrete2022},
\begin{align}
    W\left(r_1, r_2\right) = \exp{\left(-\frac{(r_1-d_1)^2}{2V_1}\right)}\exp{\left(-\frac{(r_2-d_2)^2}{2V_2}\right)}\, 
\end{align}
where,
$
    \begin{pmatrix}
    r_1 \\
    r_2
    \end{pmatrix}
    = \begin{pmatrix}
    \cos{\theta} & -\sin{\theta} \\
    \sin{\theta} & \cos{\theta}
    \end{pmatrix}
    \begin{pmatrix}
    Q \\
    P
    \end{pmatrix}
$ is the rotated coordinate system and $(d_1, d_2)$ is the displacement vector in the rotated system. The parameters $V_1$ and $V_2$ represent the variance of the distribution along the respective direction $r_1$ and $r_2$. Using the fitting procedure, the values of $V_1$ and $V_2$ are extracted for every time delay and plotted in Fig.~\ref{fig: WDF} (e). It is seen that the variances along $r_1$ and $r_2$ generalized direction modulate with the time delay between gate and probe pulses with attosecond resolution.

\subsection*{Reduction of classical noise using NRM filtering}
Quantum-noise-limited detection on an unbalanced camera is enabled by the controlled scan of the sub-cycle gate–probe delay $\tau$. The strong-field–modulated susceptibility $\chi^{(3)}(\mathcal E)$ imposes a deterministic modulation on the generated field at the optical carrier $\omega_0$ and its harmonics as a function of $\tau$~\cite{walz2025}. Consequently, the squeezed signal manifests as a set of sharp lines at delay frequencies $\Omega \in \{0, \pm\omega_0, \pm2\omega_0, \dots\}$, as shown in Fig.~\ref{fig:homodyne_signal} (d). The primary classical noise sources, including local oscillator intensity fluctuations and camera electronic noise, arise from laser and detector dynamics that are statistically independent of the interferometric delay. Their $1/f$ and drift characteristics appear along the acquisition-time axis rather than $\Omega$ (delay frequency); when projected onto the delay axis, they exhibit no $\tau$-correlation and thus form a flat, white floor in $\Omega$ without a line at $\omega_0$. Therefore, the signal and classical excess occupy distinct regions of the delay-frequency domain. NRM filtering, implemented as a low-pass at $|\Omega| \le \Omega_c$, enables synchronous (lock-in) detection of a signal at $\omega_0$ above a white noise background.

Quantum noise does not constitute a static background. Because the squeezing, or quadrature variance, is determined by $\chi^{(3)}$, quantum uncertainty modulates with $\tau$ at $\omega_0$ in tandem with the signal. The variance alternates between amplitude- and phase-squeezed as the delay is scanned, which is the central effect addressed in this work. As a result, quantum fluctuations exhibit the same $\{0, \pm\omega_0, \pm2\omega_0, \dots\}$ line structure as $S(\tau)$ and are located within the passband, on the signal lines, while the white classical residue is predominantly outside this region. A low-pass filter, rather than a band-pass at $\pm\omega_0$, is required because the DC component contains the mean-field displacement of the Wigner function, which is the dominant feature of bright squeezed light. Retaining both the DC and $\pm\omega_0$ lines preserves the displacement and the modulated quantum statistics. Since the classical residue is white in $\Omega$, the retained classical-noise variance is simply the fraction of the Nyquist span passed, $V_C'/V_C=\Omega_c/\Omega_{\mathrm{Nyq}}$. With a Nyquist of $7\omega_0$ and an effective cutoff $\Omega_c\approx1.2\,\omega_0$, only $17\%$ of the classical noise is retained, corresponding to an $83\%$ reduction or $7.5~\mathrm{dB}$ of classical-noise suppression from NRM filtering. This filter along the $\tau$-dimension is applied at each phase of the local oscillator $\phi_{LO}$ and at a single-shot index, resulting in a classical noise-filtered shot-resolved homodyne signal, as shown in Figure~\ref{fig:homodyne_signal} (b). All quantitative analyses in this work, including the tomography, the covariance and coherency matrices, the fourth-order cumulant statistics, and the Bayesian inference, are performed on the single-shot quadrature values after filtering. These values remain fluctuating random variables at each $(\tau, \phi_{LO})$. Further details of the analysis are provided in the supplementary material.

The measured squeezing is not a filtering artifact. Because quantum noise is co-located with the signal on the $\{0, \pm\omega_0\}$ lines, the passband preserves quantum noise while rejecting only the out-of-band white classical floor. Thus, the filter removes classical excess without affecting quantum information. As an additional safeguard, squeezing is reported as the ratio of two identically filtered records, coherent and squeezed. Any residual bandwidth factor $\kappa$ acting on a white noise component cancels in the ratio as $V_C'\to0$, and for the finite residue that remains, the bias is conservative. A common additive floor pulls $(\kappa V_Q^{\mathrm{sqz}}+V_C')/(\kappa V_Q^{\mathrm{coh}}+V_C')$ toward unity, so residual classical noise can only reduce the measured squeezing, never increase it. The absolute reference is established independently of the camera and the filter by the balanced-photodiode measurement in Fig.~\ref{fig: WDF} (h), where the near-coherent DFWM state at $\tau=7.6~\mathrm{fs}$ aligns with the laser shot-noise level, anchoring the coherent floor to the true quantum-noise limit.

We note that the balanced photodiode calibration is integrated over the full pulse spectrum, whereas the FR-BHD measurements are frequency-resolved. The relative intensity noise scales as 1/N, where N is the photon number. A single-frequency bin contains fewer photons than the full pulse, so the single-frequency shot noise is higher than the spectrally integrated value. The integrated reference is therefore lower than the true single-frequency shot-noise level. Squeezing, when demonstrated relative to this reference, is a conservative claim.

\subsection*{Theory}
The propagation of the quantum light field through a nonlinear medium can be described by an effective one-dimensional nonlinear Schr\"odinger equation \cite{Drummond2014}
%To simulate the quantum nature of light propagating through a nonlinear medium, we solve the following nonlinear Schr\"odinger equation.
\begin{equation}\label{NLSE}
    \frac{\partial \hat{\Psi}(z,t)}{\partial t}=\left[-\frac{\partial \omega(k_0)}{\partial k}  + \frac{i}{2}\frac{\partial^2 \omega(k_0)}{\partial k^2} + i\chi_e |\hat{\Psi}(z,t)|^2\right] \hat{\Psi}(z,t), 
\end{equation}
where $\hat{\Psi}(z,t)$ is the real-space photon field operator, $\omega(k)=\frac{k}{\sqrt{\epsilon(k) \mu_0}}$ is the dispersion relation, and $\chi_e$ is the nonlinear susceptibility of the material. 
To solve this equation, we employ the truncated Wigner approximation (TWA), in which we construct an equivalent nonlinear stochastic equation for the $\mathbf{c}$-field ($\psi$)\cite{blair:review}. We initialize the field with a coherent pulse containing $N$ photons, supplemented by vacuum noise in all modes within a plane-wave basis. The observables are obtained by averaging the simulation over 10,000 stochastic trajectories on 128 spatial grid points.
The photon field can be calculated as a function of propagation time $t$ and real-space $z$. 
%The output field then yields the multi-mode coherency matrix $(\Gamma^{(1)})_{mn} = \langle \hat{a}^{\dagger}_m \hat{a}_n\rangle$, which is calculated from the one-body density matrix,  $\rho(z,z') = \langle \hat{\Psi}^{\dagger}(z) \hat{\Psi}(z')\rangle$ \cite{book:pethik}. 
The one-body density matrix, $\rho(z,z') = \langle \hat{\Psi}^{\dagger}(z) \hat{\Psi}(z')\rangle$ \cite{book:pethik} is obtained from TWA simulations as
\begin{align}
    \rho(z,z') = \overline{\psi^*(z)\psi(z')} -\frac{1}{2}\delta(z,z')
\end{align}
where $\delta(z, z')$ is the delta function, subtracted due to the symmetrically ordered calculation of the correlation in Wigner function theory, and $\overline{\cdots}$ denotes the stochastic average. Diagonalizing the one-body density matrix
\begin{equation}
    \int dz' \rho(z,z') \phi_j(z') = \lambda_j\phi_j(z),
\end{equation}
 we obtain the eigenvalues ($\lambda_j$) and the corresponding eigenfunctions ($\phi_j(z)$) of the $j^{th}$ principal mode. The multi-mode coherency matrix $(\Gamma^{(1)})_{mn} = \langle \hat{a}^{\dagger}_m \hat{a}_n\rangle$ in the frequency basis discussed in Eq.~(\ref{eq:coherency_matrix}) of the main text is obtained by projecting the field operator onto the plane-wave modes $\varphi_m(z)$ as
 \begin{align}
     \langle\hat{a}^\dagger_m\hat{a}_n\rangle = \int dz \int d{z'} \varphi^{*}_m(z) \varphi_n(z') \langle\hat{\Psi}^{\dagger}(z)\hat{\Psi}(z')\rangle.
 \end{align}
Since the eigenspectrum of the coherency matrix is largely dominated by one principal mode, as shown in Fig.~\ref{fig:theory_plots} (f), we develop an effective single-mode description. To this end, we project onto the largely occupied principal mode, $\phi(z)$, approximating the photon field as $\hat{\Psi}(z,t) \approx \phi(z)\, \hat{a}(t)$. Its dynamics is captured by
 \begin{align}
    i \frac{d\hat{a}(t)}{dt} = \chi\, \hat{a}^\dagger(t) \hat{a}(t) \hat{a}(t). \label{eq:single_mode}
\end{align}
\begin{figure}[ht]
    \centering
    \includegraphics[width=\linewidth]{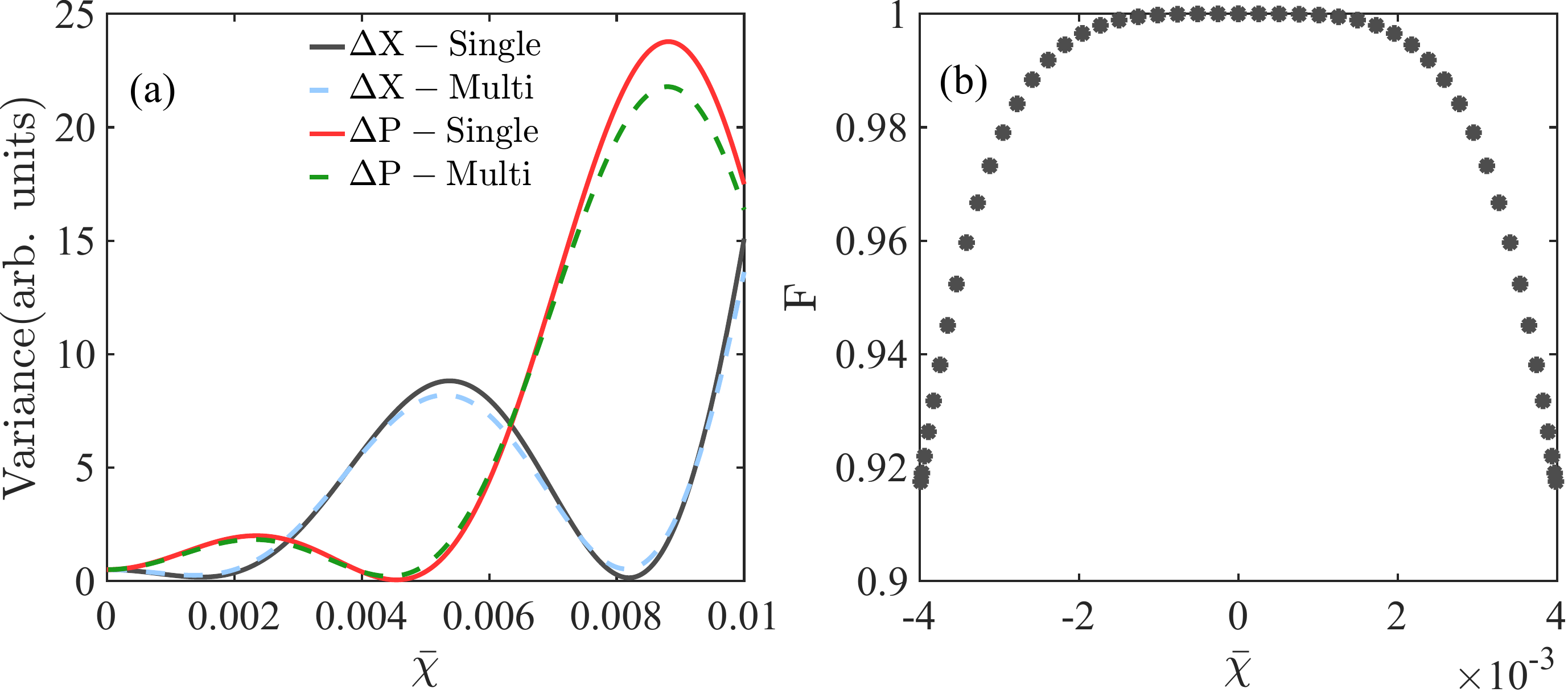}
    \caption{Variance of the quadratures as a function of the dimensionless coupling constant $\bar{\chi}$. (a) $\Delta X$ quadrature variance obtained from the multi-mode TWA (dashed blue line) with the single-mode analytical solution (solid black line). $\Delta P$ quadrature variance from the multi-mode TWA (dashed green line) with the single-mode analytical solution (solid red line). (b) Fidelity (F) of the Gaussian approximation as a function of the dimensionless parameter $\bar{\chi}$.}
    \label{fig:comparison_plot}
\end{figure}
 Here, $ \chi = \chi_e \int dz\, |\phi(z)|^4$ is the effective non-linear susceptibility for the single mode. It is directly apparent that the photon dynamics at the output only depends on $\bar{\chi} = \chi t$, the dimensionless product of nonlinearity and propagation time $t$. A comparison between the quadrature variances obtained from the full 
multi-mode TWA simulations and those from the analytically solved single-mode model is shown in Fig.~\ref{fig:comparison_plot} (a). In the considered limit of small $\bar{\chi}$, the single-mode model represents an excellent approximation to the full dynamics, which can be analytically solved. In the large mean-photon-number limit, the output quadrature variances emerging from a coherent input state are obtained from the single-mode 
model as
%  \begin{align}\label{eq;Variance}
% \Delta X &\approx \frac{1}{2}\Bigl[
%     \alpha^{*2} \, e^{\,2i\chi t}\,
%         e^{-2n\sin^{2}(2\chi t)} \, e^{\,i n\sin(4\chi t)}
%     + 
%     \alpha^{2} \, e^{-2i\chi t}\,
%         e^{-2n\sin^{2}(2\chi t)} \, e^{-i n\sin(4\chi t)}
%     + 2n + 1
% \Bigr]
% - 2\,\bigl(\Re\langle \hat{a}(t)\rangle\bigr)^{2},
% \\[1em]
% \Delta P &\approx \frac{1}{2}\Bigl[
%    -\alpha^{*2} \, e^{\,2i\chi t}\,
%         e^{-2n\sin^{2}(2\chi t)} \, e^{\,i n\sin(4\chi t)}
%    -
%    \alpha^{2} \, e^{-2i\chi t}\,
%         e^{-2n\sin^{2}(2\chi t)} \, e^{-i n\sin(4\chi t)}
%    + 2n + 1
% \Bigr]
% - 2\,\bigl(\Im\langle \hat{a}(t)\rangle\bigr)^{2}.
% \end{align}
\begin{align}\label{eq;Variance}
\Delta X &\approx \frac{1}{2}\Bigl[
    \alpha^{*2} \, e^{\,2i\bar{\chi}}\,
        e^{-2n\sin^{2}(2\bar{\chi})} \, e^{\,i n\sin(4\bar{\chi})}
    + 
    \alpha^{2} \, e^{-2i\bar{\chi}}\,
        e^{-2n\sin^{2}(2\bar{\chi})} \, e^{-i n\sin(4\bar{\chi})}
    + 2n + 1
\Bigr]
- 2\,\bigl(\Re\langle \hat{a}(\bar{\chi})\rangle\bigr)^{2},
\\[1em]
\Delta P &\approx \frac{1}{2}\Bigl[
   -\alpha^{*2} \, e^{\,2i\bar{\chi}}\,
        e^{-2n\sin^{2}(2\bar{\chi})} \, e^{\,i n\sin(4\bar{\chi})}
   -
   \alpha^{2} \, e^{-2i\bar{\chi}}\,
        e^{-2n\sin^{2}(2\bar{\chi})} \, e^{-i n\sin(4\bar{\chi})}
   + 2n + 1
\Bigr]
- 2\,\bigl(\Im\langle \hat{a}(\bar{\chi})\rangle\bigr)^{2}.
\end{align}
 Here $\alpha$ is the coherent-state amplitude and $n$ is the mean photon number. Fig.~\ref{fig:theory_plots} (e) of the main text shows these quadratures for various values of the $\bar{\chi}$. 
 %A key observation from Eq.~\ref{eq;Variance} is that the quadrature variances explicitly depend on the dimensionless parameter $\chi t$, an important feature of the Kerr dynamics. 
% Therefore, modulating the nonlinear susceptibility at a fixed propagation time directly alters the quadrature variances, as shown in Fig.~\ref{fig:theory_plots} (e). 
The output Wigner function for the single-mode model can also be solved exactly and analytically as \cite{carreno2025wigner}
%  \begin{equation}\label{eq;NonG}
% \begin{aligned}
% W(\alpha,t)
% &=\frac{2}{\pi}\,e^{-2|\alpha|^2}
% \Bigg[
% \sum_{n=0}^{\infty}
% (-1)^n\,\rho_{nn}(t)\,
% L_{n}\!\left(4|\alpha|^2\right)
% \\[4pt]
% &\qquad\qquad
% +\,2\,\Re\!\sum_{n>m}
% (-1)^m\,\rho_{mn}(t)\,
% \sqrt{\frac{m!}{n!}}\,
% (2\alpha)^{\,n-m}\,
% L_{m}^{(n-m)}\!\left(4|\alpha|^2\right)
% \Bigg].
% \end{aligned}
% \end{equation}
% cite{carreno2025wigner}
 \begin{equation}\label{eq;NonG}
\begin{aligned}
W(\alpha,\bar{\chi})
&=\frac{2}{\pi}\,e^{-2|\alpha|^2}
\Bigg[
\sum_{n=0}^{\infty}
(-1)^n\,\rho_{nn}(\bar{\chi})\,
L_{n}\!\left(4|\alpha|^2\right)
\\[4pt]
&\qquad\qquad
+\,2\,\Re\!\sum_{n>m}
(-1)^m\,\rho_{mn}(\bar{\chi})\,
\sqrt{\frac{m!}{n!}}\,
(2\alpha)^{\,n-m}\,
L_{m}^{(n-m)}\!\left(4|\alpha|^2\right)
\Bigg].
\end{aligned}
\end{equation}
Here $L^n_m$ are the associated Laguerre polynomials and 
\begin{align}
     \rho_{mn}(\bar{\chi})
    = e^{-|\alpha|^2}\frac{\alpha^m(\alpha^*)^n}{\sqrt{m!\,n!}}\,
      e^{-i\bar{\chi}\,[m(m-1)-n(n-1)]}
\end{align}
are the elements of the density matrix. The Wigner function at the output exhibits negative regions characteristic of the emergence of nonclassical light (see Fig.~\ref{fig:theory_plots}).  However, such nonclassical features appear only for large values of the 
dimensionless nonlinear parameter $\bar{\chi}$. For very small values of $\bar{\chi}$, the output state stays approximately Gaussian, and the Wigner function is well approximated by a squeezed coherent state of the form 
%Additionally, the main quadratures $X$, $P$ (use the same notation as in the main text!) can be extracted by a Gaussian fit. This approximation yields the following Wigner function}

%We can calculate the Wigner function of the photons propagating through a nonlinear medium using the following expression
\begin{equation}\label{theory_WDF}
    W(X,P)=\frac{1}{2\pi\sqrt{\text{det}(\bm\Sigma)}}\exp{\left[-\frac{1}{2}(\bm{r}-\bm{r}_0)^T\bm{\Sigma}^{-1}(\bm{r}-\bm{r}_0)\right]}\, 
\end{equation}
where $\bm{r}=(X,P)$ and $\bm{r}_0=(X_{\text{mean}},P_{\text{mean}})$. We show the fidelity ($F$) of the Gaussian state in Fig.~\ref{fig:comparison_plot} (b) by calculating the overlap between Eq.~\ref{eq;NonG} and Eq.~\ref{theory_WDF}. This indicates that the state can be well approximated by a Gaussian
distribution for small values of $\bar{\chi}$, where it is completely characterized by its amplitude and field quadratures.
% \section*{Conclusion}\label{sec6}

% We vary nonlinear response of an ultrafast DFWM interaction to achieve attosecond control of squeezing within the generated signal pulse. The range of our squeezing is from 0 to 2dB \textbf{CHECK}. This is an unprecedented degree of control over the quantum state of a pulse in the ultrafast regime. Control over squeezing gives an advantage in the pursuit of generating light with a quadrature variance more than 15dB below the quantum noise limit. We also see that control over squeezing in the ultrafast regime also adds to the bridge between ultrafast and quantum optics.

\section*{Acknowledgements}

This work was supported by the U.S. Department of Energy, Office of Science, Basic Energy Sciences, under Award $\#$DE-SC0024234 (awarded to NS). NS acknowledges support from the National Science Foundation under award $\#$2208061, which supported SP and GJE. VW acknowledges support from the U.S. Department of Energy (DOE), Office of Science, Basic Energy Sciences (BES), under Award $\#$DE-SC0026318, and by the National Science Foundation (NSF) under Award $\#$PHYS-2409630, which supported the postdoctoral researcher SKT. C.-T.L. acknowledges support from the U.S. Air Force Office of Scientific Research (AFOSR), award no. FA9550-23-1-0234. HA acknowledges the support as part of QuPIDC, an Energy Frontier Research Center, funded by the US Department of Energy (DOE), Office of Science, Basic Energy Sciences (BES), under award number DE-SC0025620.

\section*{Data Availability}
The experimental data presented in this work are publicly available at the Purdue University Research Repository (PURR). DOI: 10.4231/Z66Q-6175

\section*{Code Availability}
The data analysis codes used to analyse the experimental data are also publicly available at the Purdue University Research Repository (PURR). DOI: 10.4231/Z66Q-6175

\section*{Author contributions}
R.Z. and S.K. contributed equally to this work. N.S. conceived the experiment. R.Z constructed the experimental setup with support from E.L., F.W., and G.J.E., and performed the measurements. S.K. performed the homodyne tomography, quadrature correlation analysis and the non-Gaussianity analysis. M.E. developed and performed the Bayesian inference analysis. S.K.T and V.W. developed the theoretical model and performed the simulations. N.S. supervised the experimental work. All authors contributed to data interpretation. R.Z., S.K., S.K.T., V.W., and N.S., wrote the manuscript with input from all authors. All authors discussed the results and commented on the manuscript.

\section*{Competing interests}
The authors declare no competing interests.

% \section*{Declarations}

% Some journals require declarations to be submitted in a standardised format. Please check the Instructions for Authors of the journal to which you are submitting to see if you need to complete this section. If yes, your manuscript must contain the following sections under the heading `Declarations':


\bibliography{bibliography}% common bib file

%%%%%%%%%%%%%%%% END OF MAIN TEXT %%%%%%%%%%%%%%%

\newpage

%%%%%%%%%%%%%%%% START OF SUPPLEMENT %%%%%%%%%%%%%%%

% Figures, tables, equations and pages in the supplement are numbered S1, S2 etc.
\renewcommand{\thefigure}{S\arabic{figure}}
\renewcommand{\thetable}{S\arabic{table}}
\renewcommand{\theequation}{S\arabic{equation}}
\renewcommand{\thepage}{S\arabic{page}}
\setcounter{figure}{0}
\setcounter{table}{0}
\setcounter{equation}{0}
\setcounter{page}{1} % not 0 as \newpage already started a supplementary page
% References continue the numbering from the main text.

\begin{center}

\section*{Supplementary Materials for Attosecond control of Squeezed  Light}

Russell Zimmerman$^{1\dagger}$,
Shashank Kumar$^{1\dagger}$,
Shiva Kant Tiwari$^{2}$,
Md Ehsanuzzaman$^{1}$,
Eric Liu$^{1}$,
Francis Walz$^{1}$,
Siddhant Pandey$^{1}$,
George J. Economou II$^{1}$,
Hadiseh Alaeian$^{3,1,4}$,
Chen-Ting Liao$^{5}$,
Valentin Walther$^{2}$,
and Niranjan Shivaram$^{1,4}$

\small$^\dagger$These authors contributed equally to this work.
\end{center}

\newpage

\subsection{Nonlinear response filtering}

\begin{figure}[h]
    \centering
    \includegraphics[width=\linewidth]{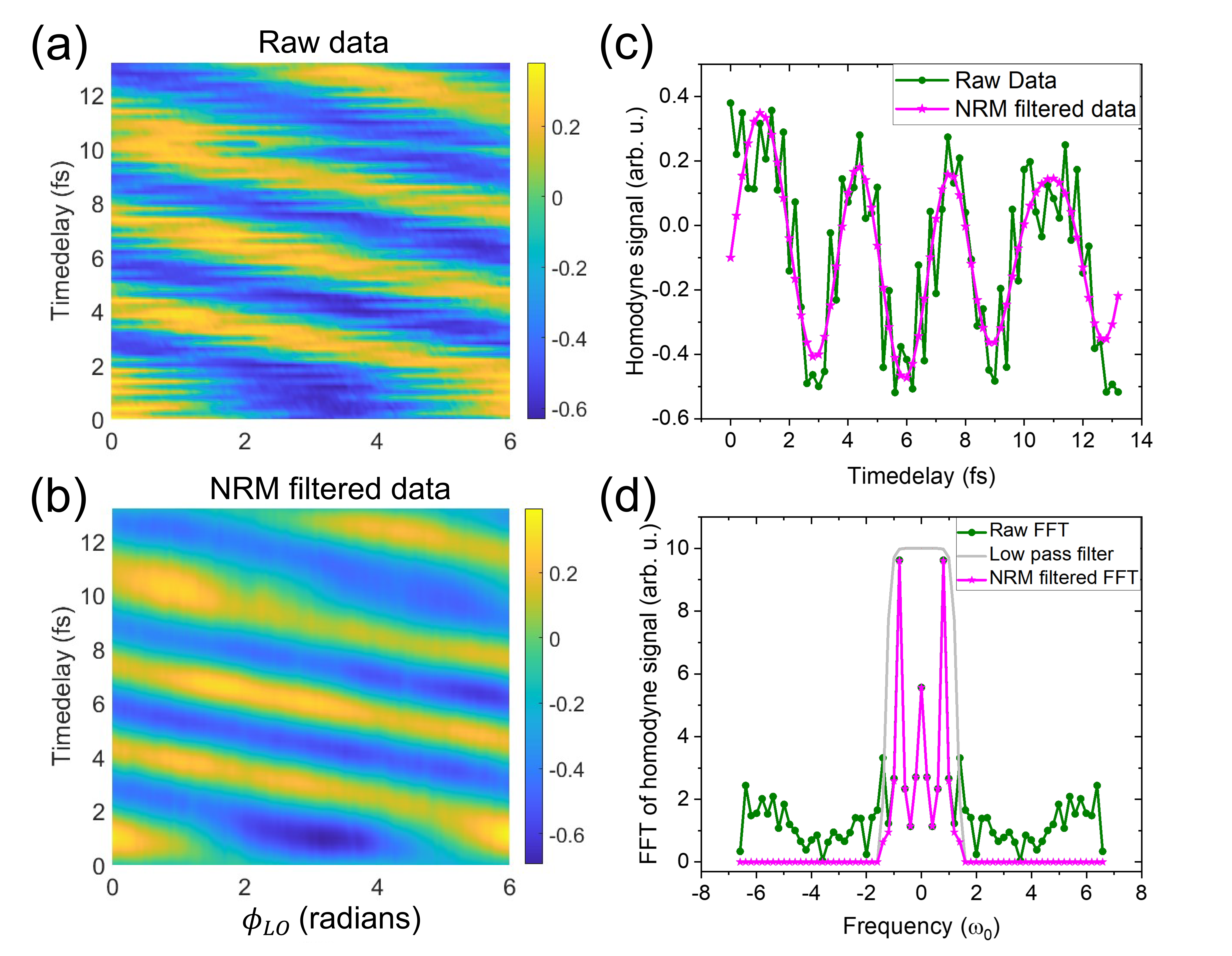}
    \caption{The averaged homodyne signal (a) raw and (b) NRM filtered data as a function of the phase of the local oscillator ($\phi_{LO}$) and time delay between gate and the probe. (c) shows the raw (green) and filtered (magenta) data as a function of time delay for $\phi_{LO}=0$, and (d) is its Fourier transform showing a $1\omega_0$ oscillation in time delay, where $\omega_0$ is the laser central frequency. We filter the raw data (green) by applying a low-pass filter (grey), which excludes all higher-order oscillation noises in the filtered signal (magenta).}
    \label{fig:homodyne_signal}
\end{figure}

Here, we present detailed plots of the homodyne signal, fully characterized as a function of the local oscillator phase $\phi_{LO}$ for various time delays. The raw data (unfiltered) are shown in Fig.~\ref{fig:homodyne_signal} (a). Figure~\ref{fig:homodyne_signal} (c) displays the signal for a specific value of $\phi_{LO}$. A Fourier transform is performed along the time-delay axis, followed by the application of a low-pass filter to retain frequency components corresponding to $\omega \leq 1\omega_0$, as illustrated in Fig.~\ref{fig:homodyne_signal} (d). The higher-order components are treated as noise. Repeating this procedure for all values of $\phi_{LO}$ yields a denoised, fully characterized homodyne signal, shown in Fig.~\ref{fig:homodyne_signal} (b).

\subsection{Quadrature and Noise correlations}

\begin{figure}
    \centering
    \includegraphics[width=\linewidth]{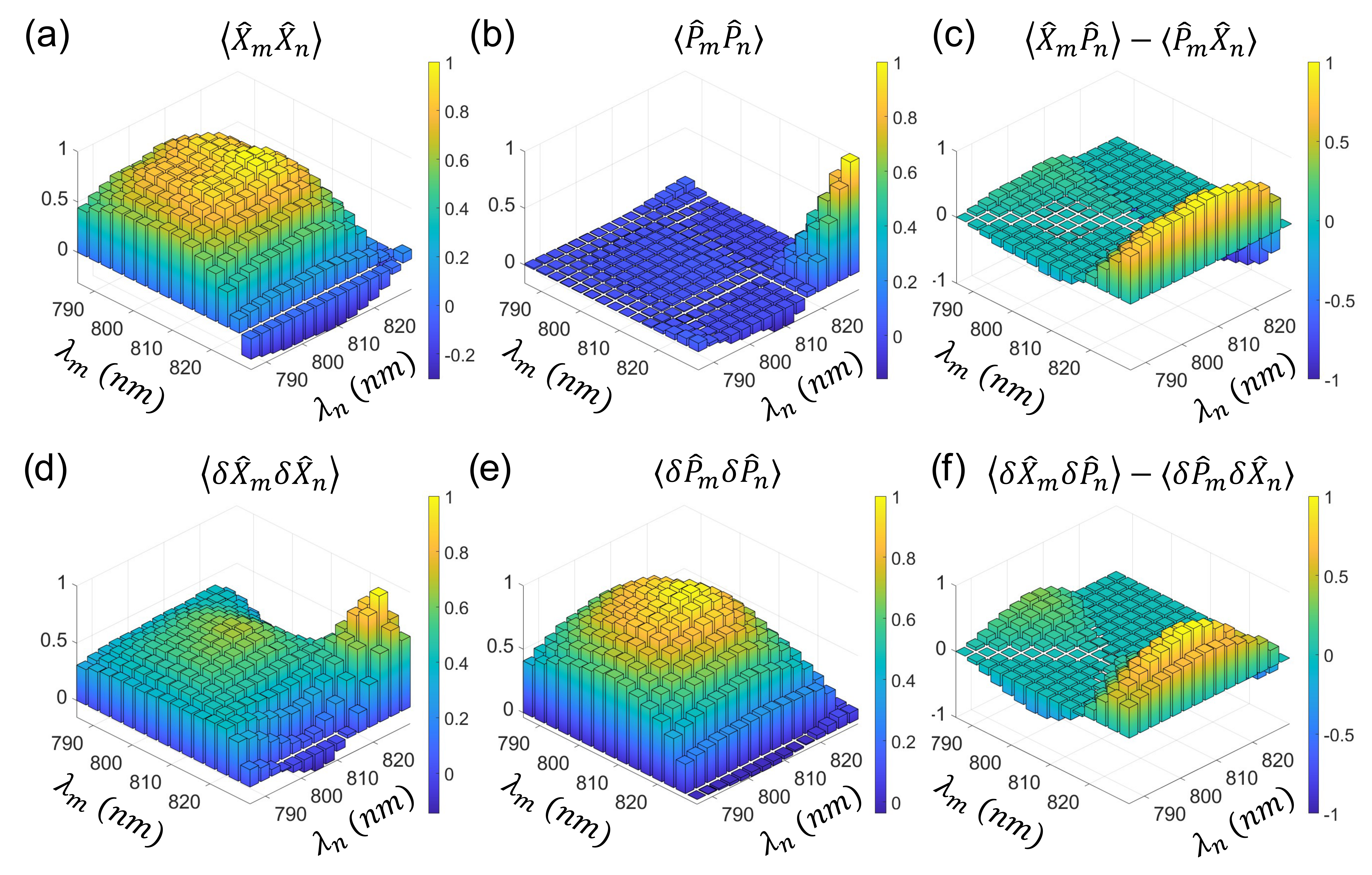}
    \caption{The elements of the covariance matrix are shown here for the amplitude squeezed state at $\tau = 10.2~\mathrm{fs}$. The diagonal terms (a) $\langle X_m X_n \rangle$ and (b) $\langle P_m P_n \rangle$ and the difference in off diagonal terms (c)$\langle X_m P_n \rangle$ - $\langle P_m X_n \rangle$. Similarly, correlation functions of quadrature noise are shown here (d) $\langle \delta X_m \delta X_n \rangle$ (e) $\langle \delta P_m \delta P_n \rangle,$ and (f) $\langle \delta X_m \delta P_n \rangle$ - $\langle \delta P_m \delta X_n \rangle$.}
    \label{fig:Quad_noise_co}
\end{figure}

As discussed in the main text, because we have the ability to measure the frequency resolved homodyne detection, it is possible to measure the quadrature correlation of the DFWM signal. The various quadrature correlations, in addition to those shown in figure 4 in the main text, are shown here in Fig.~\ref{fig:Quad_noise_co} (a) - (c). We can also extract the quantum noise correlations between different quadratures defined by $\langle \delta X_m \delta X_n \rangle$, $\langle \delta P_m \delta P_n \rangle$, $\langle \delta X_m \delta P_n \rangle$, and  $\langle \delta P_m \delta X_n \rangle$. The corresponding noise correlations are plotted in Fig.~\ref{fig:Quad_noise_co} (d) - (f).

\subsection{2D Gaussian fitting of Wigner function}
\begin{figure}
    \centering
    \includegraphics[width=\linewidth]{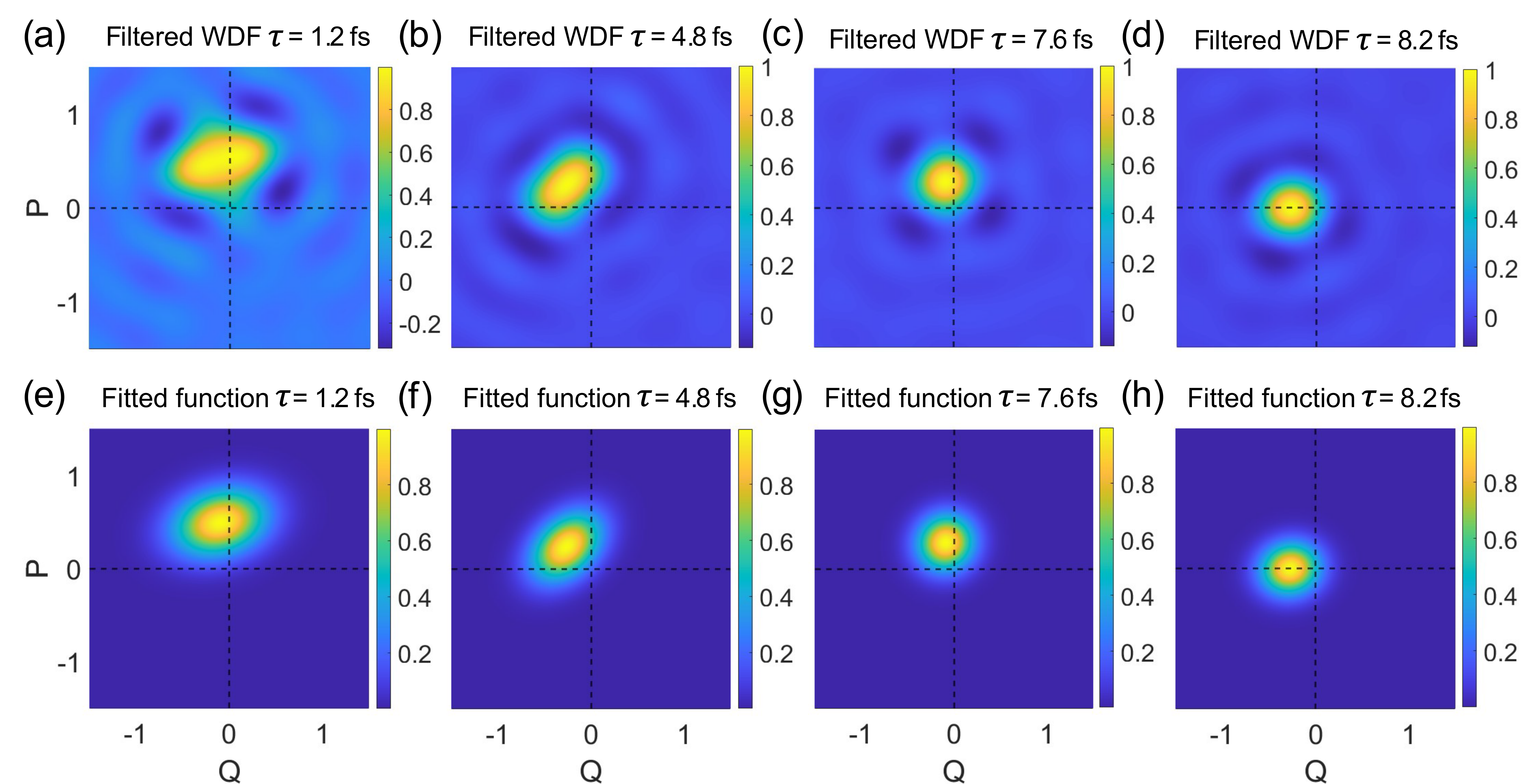}
    \caption{The experimental NRM filtered Wigner distribution function (WDF) for time delay $\tau =$ (a) $1.2~\mathrm{fs}$ (b)  $4.8~\mathrm{fs}$ (c)  $7.6~\mathrm{fs}$ and (d) $8.2~\mathrm{fs}$. The corresponding Wigner functions are fitted to a generalized 2D gaussian function shown in (e) - (h), respectively. The centroid of the Wigner distribution moves for different time delays due to a global phase of the signal with respect to the local oscillator.}
    \label{fig:filter_vs_fitted}
\end{figure}

To quantify the quadrature squeezing, we fit a generalized two-dimensional gaussian function to the measured Wigner function (see the main text for details). The measured Wigner function for $\tau = 1.2,\ 4.8,\ 7.6,\ 8.2$ fs is shown in Fig.~\ref{fig:filter_vs_fitted} (a) - (d), respectively. Their 
corresponding fitted functions are shown in Fig.~\ref{fig:filter_vs_fitted} (e) - (h), respectively.

\subsection{Variance modulation with large $\bar{\chi}$}

\begin{figure}[h]
    \centering
    \includegraphics[width=0.8\linewidth]{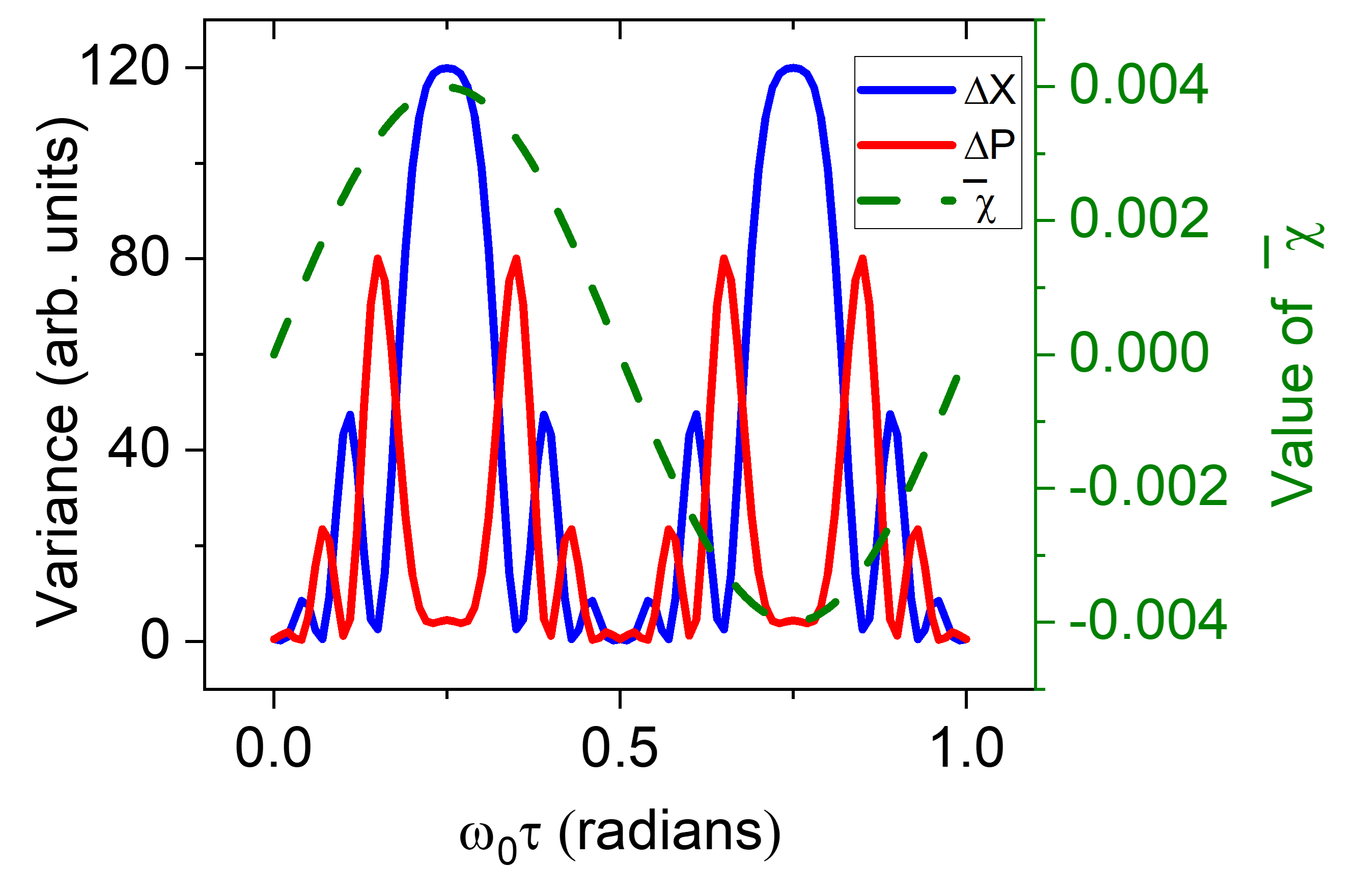}
    \caption{The oscillations of the quadrature variance $\Delta X$ and $\Delta P$ are shown for large modulation of $\bar{\chi}$. This shows that large modulation leads to introduction of higher frequencies in the variance oscillations.}
    \label{fig:supp_high_chi}
\end{figure}

Fig.~\ref{fig:theory_plots} (e) in the main text shows smooth oscillation of quadrature variance when the modulation of coupling constant $\bar{\chi}$ is small. Fig.~\ref{fig:supp_high_chi} shows that large modulation of $\bar{\chi}$ (compared to that in Fig.~\ref{fig:theory_plots} (a)-(d)) leads to more complex dynamics and introduces higher frequencies in the quadrature variance oscillations.

\subsection{Fourth-order cumulant for non-Gaussian features}
In this section, we evaluate the fourth-order cumulant of the quadrature noise $\delta X$ and $\delta P$ as an indicator of non-Gaussian states. For zero-mean variables, the fourth-order cumulant for four indices is defined as follows.
\begin{align}\label{eq:fourth_cumulant}
    \kappa_4 (A_{\alpha},B_{\beta},C_{\gamma},D_{\delta})
    &= \langle A_{\alpha}B_{\beta}C_{\gamma}D_{\delta}\rangle
       - \langle A_{\alpha}B_{\beta}\rangle \langle C_{\gamma}D_{\delta} \rangle \nonumber\\
    &\quad - \langle A_{\alpha}C_{\gamma}\rangle \langle B_{\beta}D_{\delta} \rangle
       - \langle A_{\alpha}D_{\delta}\rangle \langle B_{\beta}C_{\gamma} \rangle,
\end{align}
Here, $A,B,C,D$ each represent either the $\delta X$ or $\delta P$ quadrature with indices $\alpha, \beta, \gamma, \delta$ representing different frequency modes in the broadband pulse. For a Gaussian state, this cumulant vanishes within statistical noise, making it a valuable indicator of non-Gaussianity. However, finite sample sizes introduce statistical fluctuations that can yield a nonzero $\kappa_4$ even for a Gaussian state, potentially leading to false positives. To quantify the confidence in a nonzero $\kappa_4$, this analysis employs bootstrap resampling. Bootstrapping estimates the sampling uncertainty of a quantity by repeatedly resampling the existing data, thereby substituting for experimental repetitions that are not feasible.
 
Consider a dataset comprising $N$ measurements indexed by shot, where $\delta X = \{\delta x_1, \delta x_2, \dots, \delta x_N\}$ denotes the $X$ quadrature noise and $\delta P = \{\delta p_1, \delta p_2, \dots, \delta p_N\}$ denotes the $P$-quadrature noise, where $\delta x_i$ is the noise for $i^{\mathrm{th}}$ shot. The cumulant $\kappa_4$ is computed from these data using Eq.~\eqref{eq:fourth_cumulant}. This single estimate does not provide information about its sampling uncertainty, which is instead assessed using the nonparametric bootstrap method described below.
 
For each bootstrap replicate $b = 1, 2, \dots, M$, $N$ shot indices $\{i_1^{(b)}, \dots, i_N^{(b)}\}$ are drawn uniformly with replacement from $\{1, \dots, N\}$. The resampled dataset is then constructed by jointly selecting the corresponding entries from $\delta X$ and $\delta P$.
\begin{align}
    \delta X^{(b)} &= \{\delta x_{i_1^{(b)}}, \dots, \delta x_{i_N^{(b)}}\},
    \nonumber\\
    \delta P^{(b)} &= \{\delta p_{i_1^{(b)}}, \dots, \delta p_{i_N^{(b)}}\}.
\end{align}
 The cumulant is then recomputed on each resampled dataset to obtain the replicate $\kappa_4^{(b)}$. After $M$ replicates, the resulting set of values $\{\kappa_4^{(1)}, \kappa_4^{(2)}, \dots, \kappa_4^{(M)}\}$ approximates the sampling distribution of the cumulant. The mean and standard deviation of this set provide the bootstrap estimates of the expected value and standard error, respectively.
\begin{align}
    \bar\kappa_4 &= \frac{1}{M}\sum_{b=1}^{M}\kappa_4^{(b)},\\
    \sigma\big(\kappa_4\big)
    &= \sqrt{\frac{1}{M-1}\sum_{b=1}^{M}
        \left(\kappa_4^{(b)} - \bar\kappa_4\right)^{2}}.
\end{align}
To test the null hypothesis $\kappa_4 = 0$, which is the value expected for a Gaussian state, the standardized statistic  (z-score) is constructed as follows.
\begin{equation}\label{eq:z_score}
    z = \frac{\kappa_4 - 0}{\sigma(\kappa_4)}.
\end{equation}
Here, the numerator represents the cumulant estimated from the original (unresampled) data, while the denominator is the bootstrap standard error. This statistic expresses the measured cumulant in units of its sampling uncertainty. Under the null hypothesis and in the central limit regime, $z$ is approximately distributed as the standard normal distribution $\mathcal{N}(0,1)$. The two-sided significance for a given entry is $p = 2\left[1-\Phi(|z|)\right]$, where $\Phi$ denotes the standard normal cumulative distribution function. Entries with $|z| > 3$ (i.e., $p < 0.0027$) are classified as significantly nonzero.

The nominal rate $p(|z|>3)=0.27\%$ assumes that each $z$ is distributed as $\mathcal{N}(0,1)$ and that the entries are independent. However, these assumptions do not strictly apply in this context. Equation~\eqref{eq:fourth_cumulant} serves as a plug-in estimator and exhibits an $\mathcal{O}(1/N)$ bias for repeated-index entries (such as $XXXX$ and $XXPP$). Its finite-$N$ sampling distribution is skewed and heavy-tailed, and the entries share both shots and index symmetries. To avoid dependence on the analytic baseline, the null distribution is calibrated empirically. Multiple surrogate datasets of $N$ shots are generated from a multivariate normal distribution with covariance equal to the sample covariance $\langle \delta X\, \delta P\rangle$ of the measured quadratures. These surrogates possess vanishing cumulants above second order and thus constitute an exact Gaussian null matched to the data. Processing each surrogate through the same bootstrap cumulant $z$-score pipeline reproduces the bias, distributional shape, and correlations omitted by the $\mathcal{N}(0,1)$ approximation.
 
\begin{figure}[h]
    \centering
    \includegraphics[width=\linewidth]{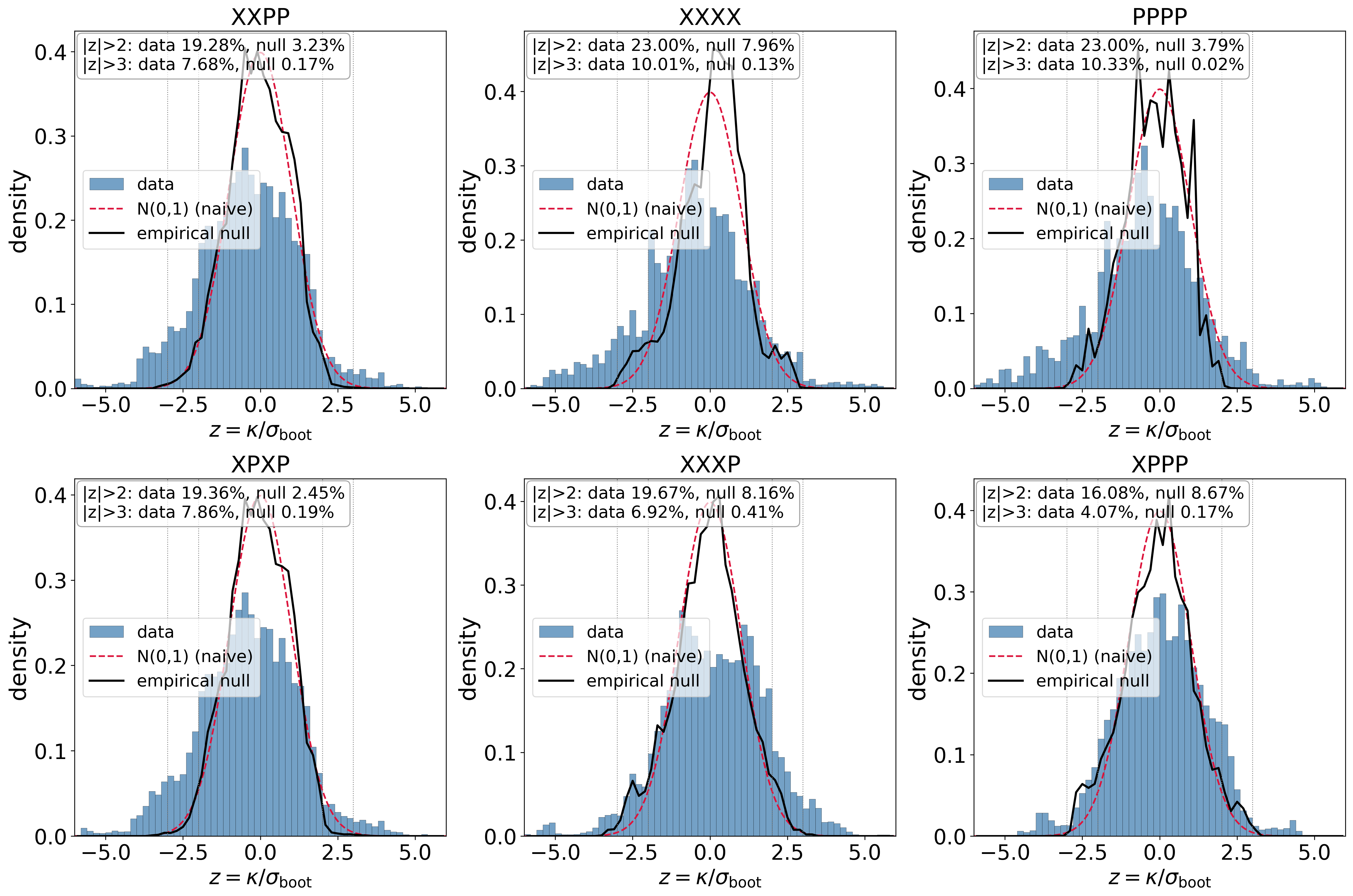}
    \caption{Histograms of the $z$-scores $z=\kappa_4/\sigma_{\mathrm{boot}}$
    (blue bins), where $\kappa_4$ is the fourth-order cumulant estimated from
    the original data and $\sigma_{\mathrm{boot}}$ is the bootstrap standard
    error. The red dashed curve is the standard normal distribution
    $\mathcal{N}(0,1)$, and the black solid curve is the empirical null
    distribution obtained from the surrogate Gaussian datasets. Each panel
    corresponds to one fourth-order quadrature combination and is annotated
    with the fractions of the measured cumulants and of the empirical null
    that fall beyond the $2\sigma$ ($|z|>2$) and $3\sigma$ ($|z|>3$)
    thresholds.}
    \label{fig:phase_sweep_hist}
\end{figure}

Figure~\ref{fig:phase_sweep_hist} presents the histogram of $z$ scores computed using Eq.~\eqref{eq:z_score} for various combinations of the $X$ and $P$ quadratures across different frequency modes and local oscillator phases. These histograms are compared to the standard normal distribution expected under the Gaussian null hypothesis. The black curve signifies the empirical null calibration calculated from surrogate Gaussian datasets. We observe that the empirical null mostly overlaps with $\mathcal{N}(0, 1)$, indicating that the plug-in estimator has very small bias. The observed $z$-score distribution deviates substantially from $\mathcal{N}(0,1)$: between $4\%$ and $10\%$ of the entries exceed $|z| > 3$, depending on the quadrature combination, whereas a Gaussian state would yield only $0.27\%$ of entries beyond this threshold by chance. This approximately 20-fold excess of supra-threshold entries cannot be explained by finite-sample noise and provides strong evidence that the measured squeezed states exhibit genuine non-Gaussianity.

\subsection{Quantifying non-Gaussianity using Standardized cumulant}\label{sec:nongauss}
Although the $z$-score presented in the preceding section demonstrates non-Gaussianity by showing that $\kappa_4$ is statistically distinguishable from zero, it does not quantify the extent of non-Gaussianity present in the state. To address this, the standardized fourth cumulant (also referred to as the non-Gaussian fraction) is employed. This metric measures the connected four-point correlation relative to the Gaussian prediction based on second-order statistics.
\begin{equation}\label{eq:standardized_cumulant}
    \tilde{\kappa}_4(\alpha,\beta,\gamma,\delta) =
    \frac{\kappa_4(\alpha,\beta,\gamma,\delta)}
         {\langle A_{\alpha}B_{\beta}\rangle \langle C_{\gamma}D_{\delta}\rangle
        + \langle A_{\alpha}C_{\gamma}\rangle \langle B_{\beta}D_{\delta}\rangle
        + \langle A_{\alpha}D_{\delta}\rangle \langle B_{\beta}C_{\gamma}\rangle}.
\end{equation}
The denominator corresponds to the disconnected (Wick) component subtracted in Eq.~\eqref{eq:fourth_cumulant}, representing the full four-point function of the reference Gaussian state. Consequently, the ratio expresses the connected correlation as a fraction of the disconnected Gaussian contribution. For a single quadrature, this reduces to $\tilde{\kappa}_4 = \kappa_4/3\kappa_2^2$, which is equivalent to one-third of the excess kurtosis.
 
\begin{figure}[h]
    \centering
    \includegraphics[width=\linewidth]{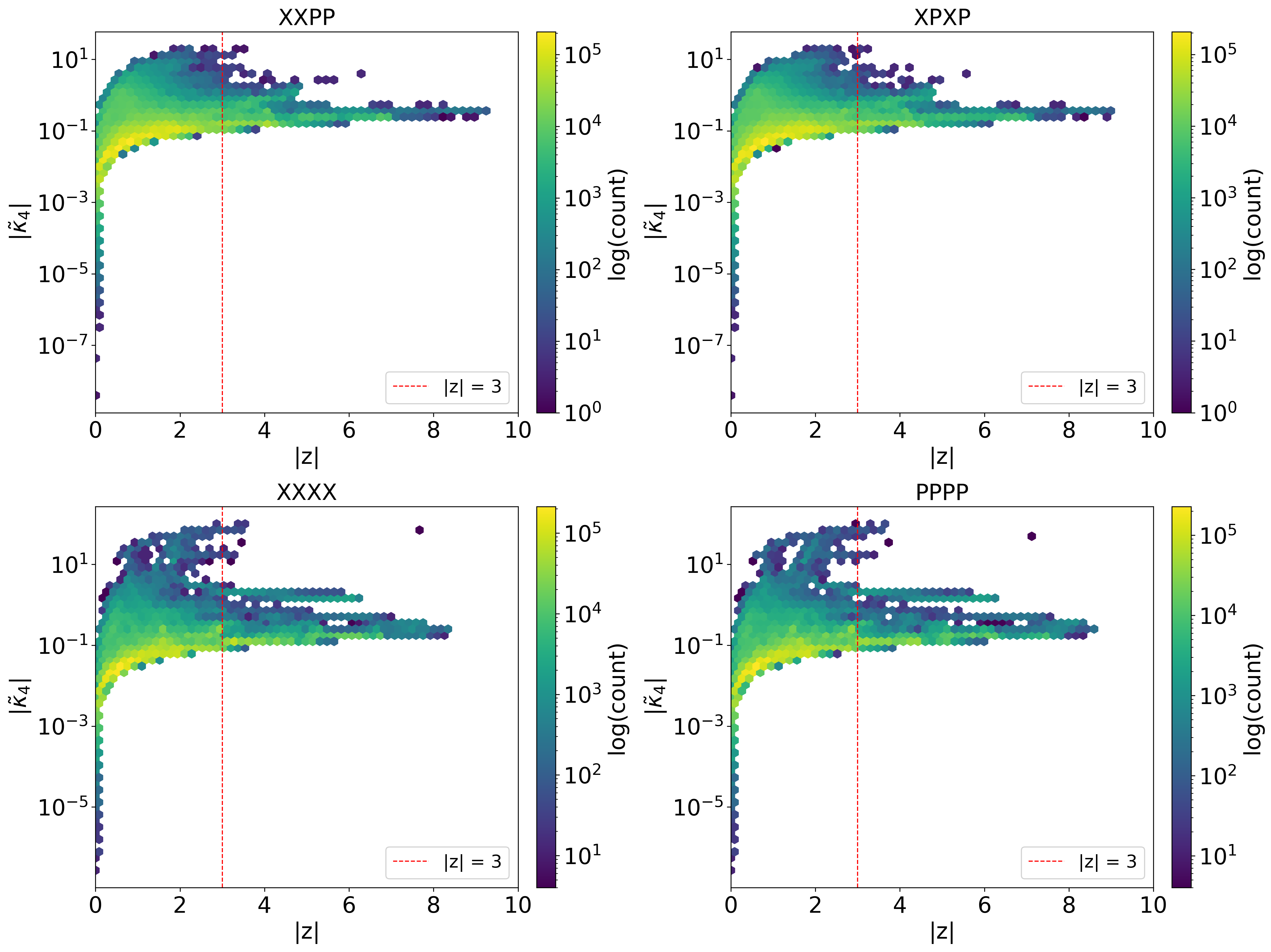}
    \caption{Joint density of the standardized fourth cumulant $\tilde{\kappa}_4$ and the $z$-score, shown as a hexbin plot. Entries with $|z|>3$ predominantly cluster near $\tilde{\kappa}_4 \approx 0.1$, indicating that the statistically significant cumulants also carry an appreciable effect size. The connected (non-Gaussian) four-point correlation thus reaches about $10\%$ of the disconnected Gaussian background for the strongly squeezed state.}
    \label{fig:phase_sweep_mag}
\end{figure}

The $z$-score (obtained via bootstrap) and the standardized cumulant $\tilde{\kappa}_4$ (calculated directly from the sample moments) are computed for the squeezed state. Figure~\ref{fig:phase_sweep_mag} presents their joint density as a hexbin plot of $\tilde{\kappa}_4$ versus $z$. Entries with $|z|>3$ cluster near $\tilde{\kappa}_4 \approx 0.1$ rather than near zero, indicating that statistically significant cumulants are associated with a substantial effect size. The connected four-point correlation therefore reaches approximately 10\% of the Gaussian background, establishing the scale of non-Gaussianity in the most strongly squeezed state at the ten-percent level.

\subsection{Reduction of classical noise using NRM filtering}
\subsubsection{Classical noise in homodyne signal detected by camera}
Let the squeezed signal field be $E_S$ and the LO be $E_{LO}=|E_{LO}|e^{i\phi_{LO}}$. After the 50/50 beam splitter in the homodyne setup the two output-port intensities are
\begin{equation}
    N_\pm \;=\; \tfrac12\big(|E_{LO}|^2+|E_S|^2\big)\;\pm\;\mathrm{Re}\!\left(E_{LO}^{*}E_S\right).
\end{equation}
A real detector pixel converts intensity to counts with a port-specific responsivity $g_j$, offset $b_j$, and additive electronic (read/dark) noise $n_j^{\mathrm{el}}$:
\begin{equation}
    I_j \;=\; g_j N_j + b_j + n_j^{\mathrm{el}},\qquad j\in\{1,2\}.
\end{equation}
where the $g_{1,2}=\bar g\,(1\pm\tfrac\epsilon2)$ is the gain of the detector, with mean $\bar g$ and mismatch $\epsilon=(g_1-g_2)/\bar g$. The homodyne difference is
\begin{align} \label{eq: homodyne_noise}
    i_- = I_1-I_2
= \underbrace{2\bar g\,\mathrm{Re}(E_{LO}^{*}E_S)}_{\text{quadrature (signal)}}
\;+\;\underbrace{\bar g\,\frac{\epsilon}{2}\,|E_{LO}|^2}_{\text{LO leakage}}
\;+\;\underbrace{\bar g\,\frac{\epsilon}{2}\,|E_S|^2+(b_1-b_2)+(n_1^{\mathrm{el}}-n_2^{\mathrm{el}})}_{\text{small / additive}}.
\end{align}
The first term is the desired quantity: $2\bar g|E_{LO}||E_S|\cos(\phi_{LO}-\theta_S)\propto X_{\phi_{LO}}$, the field quadrature amplified by the strong LO, carrying the quantum (squeezed) statistics in its fluctuations.

The second term is the bottleneck. The LO intensity fluctuates shot-to-shot through the laser relative-intensity noise (RIN): $|E_{LO}|^2=\bar I_{LO}(1+r)$, where $\bar I_{LO}$ is the average LO intensity, with $\langle r\rangle=0$, $\langle r^2\rangle=\sigma_{\mathrm{RIN}}^2$. In a balanced detector ($\epsilon\to0$) this common-mode term cancels in the subtraction. With uncorrelated camera pixels $\epsilon\neq0$ and varies pixel-to-pixel, so it does not cancel and contributes a classical fluctuation
\begin{align}\label{eq: RIN}
    \delta i_-^{\mathrm{cl}} \;=\; \bar g\,\frac{\epsilon}{2}\,\bar I_{LO}\, r \;+\;(n_1^{\mathrm{el}}-n_2^{\mathrm{el}}).
\end{align}
The shot-noise (quantum) variance of a balanced difference scales linearly with LO intensity (see first term of Eq. \ref{eq: homodyne_noise}),
\begin{equation}
    \mathrm{Var}_{\mathrm{QNL}}(i_-)\;\propto\;\bar g^2\,\bar I_{LO},
\end{equation}
whereas the RIN leakage (Eq. \ref{eq: RIN}) scales quadratically,
\begin{equation}
    \mathrm{Var}_{\mathrm{cl}}(i_-)\;=\;\Big(\bar g\,\tfrac{\epsilon}{2}\,\bar I_{LO}\Big)^2\sigma_{\mathrm{RIN}}^2+\sigma_{\mathrm{el}}^2 .
\end{equation}
Therefore the ratio,
\begin{equation}\label{eq:variance_ratio}
    \;\frac{\mathrm{Var}_{\mathrm{cl}}}{\mathrm{Var}_{\mathrm{QNL}}}\;\sim\;\frac{\epsilon^2\,\sigma_{\mathrm{RIN}}^2}{4}\,\bar I_{LO}\;+\;\frac{\sigma_{\mathrm{el}}^2}{\bar g^2\bar I_{LO}}\;
\end{equation}
grows with LO intensity. A strong LO (here a 40:1 intensity ratio) is exactly what one needs to lift the quadrature signal above electronic noise (second term in Eq. \ref{eq:variance_ratio}), but it simultaneously amplifies the gain-mismatch leakage (first term in Eq. \ref{eq:variance_ratio}). There is no LO power at which both terms are negligible. It is the leakage $\propto\epsilon^2\bar I_{LO}$, not a deficiency of the quantum source, that sets the floor noise. Averaging over shots reduces the uncertainty of a variance estimate. It cannot remove an excess variance that is physically present in every shot. Removing the excess requires spectral separation, which the NRM modulation provides.

\subsubsection{NRM encodes the quantum signal in the delay variable}
The main feature of this experiment, absent in conventional homodyne, is the controlled scan of the sub-cycle gate–probe delay $\tau$. It has been shown both experimentally and from a microscopic model, that in the presence of a moderately strong field the third-order susceptibility can be modeled an instantaneous function of the driving field, $\chi^{(3)}\!\to\!\chi^{(3)}(\mathcal E)$, expandable as \cite{walz2025}
\begin{equation}
    \chi^{(3)}(\mathcal E)=\chi^{(3)}_0+a\,\mathcal E+b\,\mathcal E^2+\mathcal O(\mathcal E^3).
\end{equation}
This is not the same as including higher orders in perturbation (via $\chi^{(5)}$, $\chi^{(7)}$, etc) but is a model to incorporate non-perturbative effects (field-driven band structure modulation) in an intuitive, perturbative picture \cite{walz2025}. Inserting the three pulses with a delay $\tau$ between the phase-locked gates and the probe ($E_{1,2}(\omega)e^{-i\omega\tau}$, $E_3(\omega)$) into the effective DFWM polarization yields \cite{walz2025}
\begin{equation}
    P^{(3)}_{\mathrm{eff}}=f_0(\omega)+f_1(\omega)e^{-i\omega_0\tau}+g_1(\omega)e^{+i\omega_0\tau}+f_2(\omega)e^{-2i\omega_0\tau}+\dots,
\end{equation}
i.e. the generated signal field's amplitude and phase oscillate as $\sum_n c_n e^{-in\omega_0\tau}$. The time-integrated amplitude is dominated by the $1\omega_0$ component, with the chirp carrying $2\omega_0$ and higher harmonics, in agreement with both time-dependent perturbation theory (with field-dressed bands) and the semiconductor Bloch equations.

Consequently, the generated squeezed field is a deterministic, periodic function of $\tau$ at the optical carrier $\omega_0$. Its quadrature, read out by the homodyne term, varies as
\begin{equation}
    S(\tau,\phi_{LO})=2\bar g\,|E_{LO}|\,|E_S(\tau)|\cos\!\big(\phi_{LO}-\theta_S(\tau)\big),
\qquad
|E_S(\tau)|,\theta_S(\tau)\ \text{modulated at}\ \{0,\omega_0,2\omega_0,\dots\}
\end{equation}
where the delay-frequency $\Omega$ as the Fourier conjugate of $\tau$, the signal's spectral support is the discrete set $\Omega\in\{0,\pm\omega_0,\pm2\omega_0,\dots\}$, which is a line spectrum, sharply peaked, exactly as observed in Fig. \ref{fig:homodyne_signal} (d), where the retained content sits in a narrow band at $\pm1\omega_0$. The crucial point is that the quantum object of interest, which is the squeezed mode whose quadrature variance we aim to measure, is modulated at this $\omega_0$ frequency in time delay. The net input field modulation inside the nonlinear medium due to sub-cycle time delay modulates the quantum field itself and not merely a classical envelope.

\subsubsection{NRM spectral separation: the classical residue is incoherent with time delay $\tau$}
Two of the main sources of classical noise introduced are shot-to-shot intensity fluctuations of the LO and the electronic noise of the camera. However, there is no physical pathway by which stepping the delay stage in attosecond increments imprints an $\omega_0$-periodic structure onto the laser's intensity fluctuations. Time delay is set by an interferometric delay line, whereas the noises are set by gain/pump dynamics of the oscillator and amplifier. The two are statistically independent.

Laser drift and $1/f$ noise are real, but they live on the acquisition-time axis (real time over which frames are collected), not on the delay-frequency axis $\Omega$. Because the classical noise is statistically independent of the delay parameter $\tau$, its projection onto the $\tau$-axis has no $\tau$-correlation, i.e. a flat (white) power spectrum in $\Omega$, with no rise toward $\Omega=0$ and no line at $\Omega=\omega_0$. The $1/f$ structure exists, but along a different axis that the $\tau$-FFT does not resolve. So in the $\Omega$ domain the signal is a set of sharp lines and the classical excess is a uniform floor underneath them:
\begin{equation}
    \tilde S(\Omega)\ \text{= sharp lines at}\ \{0,\pm\omega_0,\pm2\omega_0,\dots\},
\qquad
\widetilde{\delta i^{\mathrm{cl}}}(\Omega)\ \text{= flat (white), no $\omega_0$ line.}
\end{equation}
This is precisely the situation exploited by lock-in detection and by optical chopping: a signal is impressed on a carrier at a chosen frequency, and a narrow band selection around that carrier recovers the signal while rejecting the out-of-band noise. The NRM is the optical modulator, the carrier frequency is fixed by the medium at $\omega_0$ and the low-pass to $|\Omega|\le\Omega_c$ followed by inverse fourier transform is the band selection. Because the noise floor is white, a low-pass that passes a fraction $\Omega_c/\Omega_{\mathrm{Nyq}}$ of the Nyquist span retains exactly that fraction of the total classical-noise variance and discards the rest.

\subsubsection{Quantitative estimate of the classical-noise reduction}
The low-pass is retained (not a band-pass) for a physical reason: the DC component in $\Omega$ carries the mean DFWM field, i.e. the displacement of the Wigner function from the origin. For bright squeezed light this displacement is the dominant feature of the state, so the filter must keep $\Omega=0$. A strict band-pass at $\pm\omega_0$ would recenter the reconstructed WDF at the origin and yield the wrong state. The low-pass keeps DC and the $\pm\omega_0$ signal lines together. The cost is that it also retains the noise-only bins between DC and the cutoff that retained classical noise.

Because the classical residue is white in $\Omega$, the retained fraction of its variance is just the fraction of the Nyquist span passed:
\begin{align}
    \frac{V_C'}{V_C}=\frac{\text{passband bins}}{\text{total bins}}=\frac{\Omega_c}{\Omega_{\mathrm{Nyq}}} .
\end{align}

With the 808 nm cycle $T_0=\lambda/c=2.69$ fs and Nyquist $\Omega_{\mathrm{Nyq}}=7\omega_0$, the delay step is $\Delta\tau=T_0/(2\cdot7)=192$ as (consistent with the $\sim 200$ as stage resolution). A $\tau$-record of $\approx13.4$ fs gives $N\approx70$ bins over $\pm7\omega_0$, i.e. $\Delta\Omega=14\omega_0/70=0.2\,\omega_0$ per bin. From Figure \ref{fig:homodyne_signal} (d) the low-pass flat-top with roll-off has an effective cutoff $\Omega_c\approx1.2\,\omega_0$. Hence
\begin{equation}\label{eq: classical_noise}
    \frac{V_C'}{V_C}\approx\frac{1.2\,\omega_0}{7\,\omega_0}\approx0.17
\end{equation}
This means that 83\% of the classical noise has been removed and 17\% of the noise is retained which means $7.5~\mathrm{dB}$ of classical noise reduction from NRM filtering.

\subsubsection{Why measured squeezing is not filtering artifact}
The per-shot homodyne signal at fixed $\phi_{LO}$ can be decomposed into three pieces:
\begin{equation}
    i_-(\tau)=\underbrace{S(\tau)}_{\text{lines at }n\omega_0}
+\underbrace{\delta q(\tau)}_{\text{quantum}}
+\underbrace{\delta c(\tau)}_{\text{classical, incoherent in }\Omega},
\end{equation}
with $\langle\delta q\rangle=\langle\delta c\rangle=0$, $\mathrm{Var}(\delta q)=V_Q$ (the quantity of physical interest, which is shot-noise-limited for a coherent state, sub-shot-noise for a squeezed state), and $\mathrm{Var}(\delta c)=V_C$.

A linear filter with transfer function $H(\Omega)$ acts identically on all three. As the signal and quantum noise periodically depends on time delay $\tau$ (because the DFWM signal modulates), the surviving variance is set by the filter's noise-equivalent bandwidth:
\begin{equation}
    V_Q \;\longrightarrow\; \kappa\,V_Q,\qquad \kappa=\frac{\int |H(\Omega)|^2\,d\Omega}{\int d\Omega}\ \le 1 .
\end{equation}
For the excess classical noise, the low-pass rejects the fraction of $V_C$ lying above the cutoff: $V_C\to V_C'=(\Omega_c/\Omega_{\mathrm{Nyq}})\,V_C\approx0.17\,V_C$ (see Eq. \ref{eq: classical_noise}). The retained residue is small ($\sim7.5$ dB below the signal) but not exactly zero; the ratio argument below is what makes the result robust to whatever residue remains, provided it is common to the coherent and squeezed states.

The squeezing is reported as a ratio of two identically filtered signals. Let the coherent reference (near-$\tau$ where the two quadrature variances are nearly equal i.e $V_1\!\approx\!V_2$) and the squeezed state be passed through the $H(\Omega)$:
\begin{equation}
    V^{\mathrm{filt}}_{\mathrm{coh}}=\kappa\,V_Q^{\mathrm{coh}}+V_C'^{\,\mathrm{coh}},
\qquad
V^{\mathrm{filt}}_{\mathrm{sqz}}=\kappa\,V_Q^{\mathrm{sqz}}+V_C'^{\,\mathrm{sqz}} .
\end{equation}
If the in-band classical residues are small (~7.5 dB below signal) and common to both records, the squeezing is
\begin{equation}
    \frac{V^{\mathrm{filt}}_{\mathrm{sqz}}}{V^{\mathrm{filt}}_{\mathrm{coh}}}
=\frac{\kappa\,V_Q^{\mathrm{sqz}}}{\kappa\,V_Q^{\mathrm{coh}}}
=\frac{V_Q^{\mathrm{sqz}}}{V_Q^{\mathrm{coh}}}
\end{equation}
when $V_C'\to0$. The measured squeezing is therefore not a filtering artifact: it is the ratio of the genuine quantum variance of the squeezed state to that of the coherent state. Discarding bandwidth in the NRM filtering process lowers both floors by the same factor and leaves their ratio untouched.

For the finite residue quantified in the previous section, the cancellation is not exact, but the bias goes in the favorable direction: a common additive $V_C'>0$ in both numerator and denominator pulls the ratio $(\kappa V_Q^{\mathrm{sqz}}+V_C')/(\kappa V_Q^{\mathrm{coh}}+V_C')$ toward unity, i.e. toward less apparent squeezing. A residual classical floor can only reduce the measured squeezing, never increase it, so the filter cannot manufacture a sub-shot-noise result.

It is important to fix the absolute reference, i.e. to prove that $V_Q^{\mathrm{coh}}$ truly equals the shot-noise level rather than some filtered-down pseudo-floor. This is measured independently of the camera and of the filter and shown in Figure \ref{fig: WDF} (h). On a standard balanced-photodiode detector (where $\epsilon\!\approx\!0$ and no NRM filtering is applied), the DFWM near-coherent state at $\tau=7.6$ fs sits on the laser shot-noise level. That measurement demonstrated that the coherent state is at the true shot-noise limit.

\subsection{Bayesian Inference}
From conditional probability, we can write,
\begin{equation}
P\!\left(\frac{\mathrm{Squeezed}}{\mathrm{Data}}\right)
= \frac{P\!\left(\frac{\mathrm{Data}}{\mathrm{params}}\right) P(\mathrm{params})}
       {P(\mathrm{Data})}.
\label{eq:bayes}
\end{equation}
The left hand side of Eq. \ref{eq:bayes} tells us that given a set of data, what is the probability that these data represent a squeezed state. The right hand side of Eq. \ref{eq:bayes} has three terms. $P\!\left(\frac{\mathrm{Data}}{\mathrm{params}}\right)$ tells us the probability of getting our data from a squeezed state. $P(\mathrm{params})$ is called the prior, which represents the probability distribution of a particular parameter from a squeezed state. $P(\mathrm{Data})$ acts as a normalization factor for a fixed dataset. This whole process of getting a probability distribution, also called posterior, is known as Bayesian inference.
Applying Bayesian inference gives us the following expression:

\begin{equation}
\underbrace{P(\lambda \mid \mathrm{Data})}_{\text{posterior}}
\;\propto\;
\underbrace{\prod_{j=1}^{n_\theta}\prod_{k=1}^{N}
   p\!\left(m_{jk}\mid \lambda,\theta_j\right)}_{\text{likelihood}}\;
\underbrace{P(\lambda)}_{\text{prior}} ,
\label{posterior}\end{equation}

where $\lambda$ is the complete set of parameters that describes the state of squeezed and coherent states,
\begin{equation}
\lambda = \{\, g,\ \bar X,\ \bar P,\ \nu,\ r,\ \phi,\ \bar X_{\mathrm{ref}},\ \bar P_{\mathrm{ref}} \,\}.
\end{equation}
$m_{jk}$ is the N (1000) shots data from the normalized homodyne ($i_{-}$) dataset for a fixed $\tau$ delay and fixed local oscillator phase; $\theta_j$ is the corresponding local oscillator phase.
Here, $g$ is the gain factor of the camera, and $(\bar{X},\bar{P})$ represents the center of the elliptical Wigner Distribution Function (WDF) in the phase space. We get $(\nu,r,\phi)$ by writing the Covariance matrix in the following form: $\Sigma = \nu\,R(\phi)\,\mathrm{diag}(e^{-2r},e^{+2r})\,R(\phi)^{\!\top}$. And $(\bar X_{\mathrm{ref}},\ \bar P_{\mathrm{ref}})$ represents the center of the distribution of the coherent state in phase space. Equation \ref{posterior} gives us a confidence level of our data representing a squeezed state. This procedure does not eliminate the need for a shot-noise reference and requires an independent shot-noise calibration, which we have performed using our balanced photodiode measurement. 

Each shot is drawn from a normal distribution whose mean and variance the model
fixes,
\begin{equation}
m_{jk}\sim \mathcal{N}\!\left(g\,\mu_{\theta_j},\ g^2 V_{\theta_j}\right),
\qquad
p\!\left(m_{jk}\mid\lambda,\theta_j\right)
= \frac{1}{\sqrt{2\pi g^2 V_{\theta_j}}}\,
  \exp\!\left[-\frac{\left(m_{jk}-g\,\mu_{\theta_j}\right)^2}{2 g^2 V_{\theta_j}}\right],
\label{eq:gauss}
\end{equation}

The empirical mean of each $(\tau, \phi_{LO})$ cell is subtracted from the shots before the inference. The likelihood therefore acts on the zero-mean fluctuations. Here $\mu_{\theta_j}$ is the mean and $V_{\theta_j}$ is the variance of the distribution of the normalized homodyne ($i_{-}$) data for a fixed local oscillator phase $\theta_j$.
Because the deviation from a Gaussian state is small as shown in section~\ref{sec:nongauss} we can make the assumption of a normal distribution here.
The total prior of the equation is the product of priors related to each parameter. Priors are logical distributions that represent our assumptions on the state containing some free parameters that we can choose.
Fixing these parameters and their priors, we scan over the whole parameter space of squeezed state, and we get Posteriors. We obtain a distribution of squeezing values from the posterior samples. We apply a Markov Chain Monte Carlo (MCMC) procedure to scan over the parameter space. It is not computationally convenient to use the equation \ref{posterior} directly. The following equation is much more convenient:

\begin{equation}
\ln P(\lambda \mid \mathrm{Data}) =\ln P(\mathrm{Data}\mid\lambda)+\ln P(\lambda)
= \sum_{j=1}^{n_\theta}\sum_{k=1}^{N}
  \left[
    -\frac{1}{2}\ln\!\left(2\pi g^2 V_{\theta_j}\right)
    -\frac{\left(m_{jk}-g\,\mu_{\theta_j}\right)^2}{2 g^2 V_{\theta_j}}
  \right] + \sum_{k=1}^8 lnP(\lambda_k).
\label{eq:lnL}
\end{equation}
Gaussian distribution as our priors is a good choice, as the central limit theorem plays a role in deciding this. The prior on the shot noise level $\sigma^2_{SNL}$ (which encodes $g$) is derived from the variance of the coherent reference state measured at the same detector.
The distribution of $\phi$ is assumed to be uniform across $[0, \pi)$. The remaining parameters satisfy physical constraints imposed by quantum mechanics. For the state to be squeezed, we need $r>0$; therefore, a half-normal distribution for $r$ is logical. From the uncertainty principle $\text{Det}(\Sigma)\geq 1/4$, this relation gives us a bound on the parameter $\nu$, also called the symplectic eigenvalue. Thus, $\nu \geq 1/2$. For computational purpose, we define $\nu = (1/2)+\bar n_{th}$, where $\bar n_{th} \geq 0$. Therefore, a half-Gaussian distribution for this new parameter $\bar n_{th}$ is assumed.

If $g$ is the gain factor of the camera, then for a coherent state, we get,
\begin{equation}
    \sigma^2_{SNL} = \frac{g^2}{2}
\end{equation}

 Therefore, calculating the distribution of $\sigma^2_{SNL}$ from the data directly gives us the prior for $g$.

% \begin{center}
%     \begin{figure}[h]
%         \includegraphics[width=\linewidth]{Bayesian and Surrogate figures/prior_all.png}
%     \caption{Distribution of all the priors assumed in order to apply Bayesian Inference}
% \end{figure}
% \end{center}
 
 For the MCMC algorithm, in the parameter space, imposing all the constraints for the state to be squeezed, we deploy an ensemble of walkers using an affine-invariant stretch-move sampler. At each step, proposed moves are accepted or rejected according to the Metropolis-Hastings criterion applied to the log-posterior.

\begin{figure}[htbp]
    \centering

    \includegraphics[width=\linewidth]{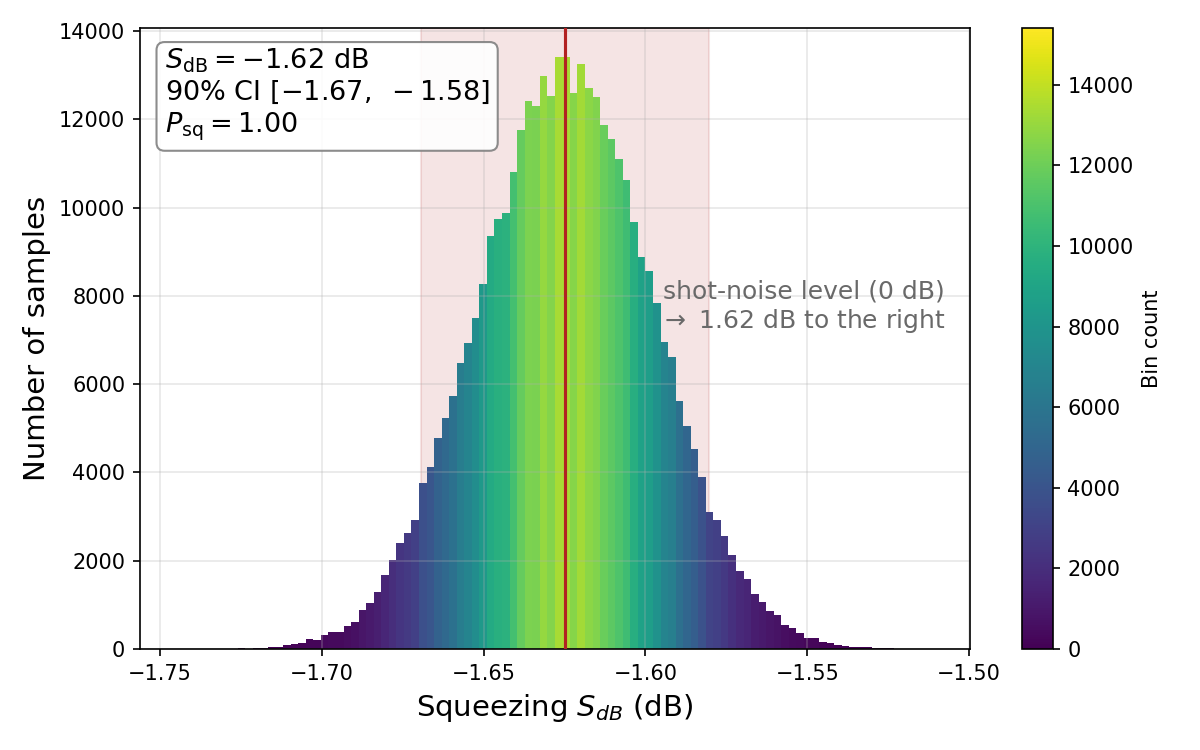}

    \includegraphics[scale=0.32]{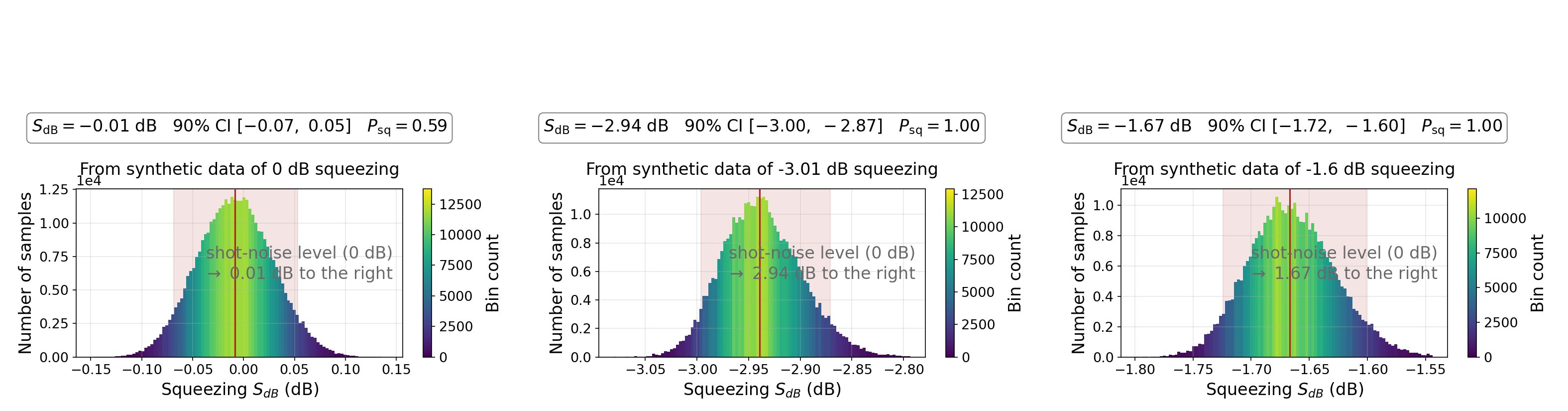}
    
    \caption{(a) Top figure is the marginal posterior distribution of the squeezing $S_{dB}$, computed from 384,000 post-burn-in samples of the affine-invariant ensemble sampler from experimental dataset. The red line marks the mean. (b) Bottom figures are the marginal posterior distribution from a synthetic dataset with known squeezing.}
    %\label{fig:combined_images}
\end{figure}

From the posterior samples, we identify those for which the squeezed quadrature variance falls below the shot noise level. The analysis is performed at the wavelength bin corresponding to $808 \;nm$ at $\tau = 10.2\; fs$, the delay of maximum squeezing. The sampler used $64$ walkers and $8000$ steps with $2000$ burn-in steps discarded, giving $384,000$ posterior samples. For our dataset, we find that all the points have squeezing below the shot noise level, and $P_{sq}=1.0$ indicates that all posterior samples yield a squeezed quadrature variance below the shot noise level. $90\% $ of the elements of the squeezing set are between $[-1.67,-1.58]\; dB$ and the mean is $S_{dB} = -1.62\; dB$.

The result of this method applied to a synthetic dataset that represents a state of known squeezing shows its effectiveness. The squeezed state from the synthetic dataset had the same variance $V_0$ (it represents the role of $\nu$ described earlier, which is fixed, $\nu = 1/2$) as the coherent state of our experiment, and different squeezing was generated using the following relation,
\begin{equation}
    V_\theta = V_0 (e^{-2r} \cos^2(\theta - \phi)  +e^{2r} \sin^2(\theta - \phi) )
\end{equation}
Here, $r$ and $\phi$ represent the parameters, and $\theta$ is the local oscillator phase.
We generated three synthetic datasets with programmed squeezing of 0 dB, -3.01 dB, and -1.62 dB relative to the shot noise level.

% \begin{center}
%     \includegraphics[scale=0.32]{Bayesian and Surrogate figures/histogram_combined.png}
% \end{center}

The result shows that Bayesian inference determines the squeezing with certain accuracy. Here it must be mentioned that for 0 dB squeezed data $P_{sq}=0.59$, which is expected, as theoretically 50\% of the $S_{dB}$ should have less than zero squeezing in the distribution and the mean should be 0 dB. For other two datasets, the recovered squeezing is $-2.94$ dB 90\%CI $[-3.00,-2.87]$ for the injected $-3.01$ dB squeezing and mean $-1.67$ dB for the injected $-1.62$ dB case. For the $-3.01$ dB injected case, the squeezing lies outside of the interval, with a minor deviation of 0.1 dB. Overall, the method recovers the injected squeezing level to within the statistical uncertainty across all three test cases.

\subsection{Simulated data through NRM filter}
NRM filter acts as a linear operator on the dataset. Linear operators acting on the dataset do not introduce non-Gaussianity in the Gaussian distribution. In this section, we test the effect of the NRM filter on a synthetic dataset and show that it does not artificially introduce a deviation from Gaussianity. We generate a dataset very similar to our experimental data but having a Gaussian distribution with some classical noise. We then apply our NRM filter and test the Gaussianity of the filtered dataset.

For this analysis, we use the definition of z-score described in equation \ref{eq:z_score}. Let $z$ denote the measured z-score for a given quadrature combination. With only 1000 shots, $\kappa_4$ computed from real data will never be exactly zero even for perfectly Gaussian light. There is always some scatter from having a finite sample size. The question is whether the scatter we see is larger than that ordinary sampling scatter, or whether the NRM filter introduces deviation from the expected distribution for a Gaussian. To answer this, we generate a batch of synthetic data that we know for certain is Gaussian. We generate it from Gaussian random numbers with the same covariance as in the experiment, and run it through the exact same NRM filter and the exact same $\kappa_4$ calculation used on the real data. 
The method of generating this dataset starts with the covariance matrix $\Sigma_{\text{joint}}$,
 \begin{equation}
\Sigma_{\text{joint}} =
\begin{pmatrix}\Sigma_{XX} & \Sigma_{XP}\\ \Sigma_{XP}^\top & \Sigma_{PP}\end{pmatrix} .
\end{equation}

Here we used the fact that $\langle P_mX_n\rangle = \langle X_nP_m\rangle$ for real numbers.
We can construct a vector $\mathbf{x}$ with a given covariance matrix $\Sigma$ by multiplication of a matrix $L$ with a vector of independent standard Gaussian numbers $\mathbf{z}$,
\begin{equation}
\mathbf{x} = L\mathbf{z}, \qquad \mathbf{z}\sim\mathcal N(0,I).
\end{equation}
Since $\text{Cov}(\mathbf{z}) = I$, the covariance of $\mathbf{x}$ is
\begin{equation}
\text{Cov}(\mathbf{x}) = L\,\text{Cov}(\mathbf{z})\,L^\top = LL^\top .
\label{eq:LLT}
\end{equation}
So the entire construction reduces to finding any matrix $L$ satisfying $LL^\top = \Sigma$, a matrix square root of $\Sigma$. Because $\mathbf{x}$ is a fixed linear combination of independent Gaussian numbers, it is Gaussian by construction, for any choice of $L$ satisfying Eq.~\eqref{eq:LLT}.
 
One convenient choice of $L$ comes from the eigenvalue decomposition of the real measured covariance matrix $\Sigma = V\Lambda V^\top$, where $V$ contains the eigenvectors and $\Lambda$ the (non-negative, for a genuine covariance matrix) eigenvalues along its diagonal. Taking
\begin{equation}
L = V\sqrt{\Lambda}
\end{equation}
gives $LL^\top = V\sqrt{\Lambda}\sqrt{\Lambda}^\top V^\top = V\Lambda
V^\top = \Sigma$ exactly, satisfying Eq.~\eqref{eq:LLT} by construction.
By constructing this matrix $L$, we generate the data $\mathbf{x}$ that is Gaussian as described above. Since we built the input to be Gaussian, any corresponding $\kappa_4$ is, by definition, pure processing noise, and that gives us a reference for how big $\kappa_4$ can get for a purely Gaussian input.

\begin{center}
    \begin{figure}[t]
    \includegraphics[width=\linewidth]{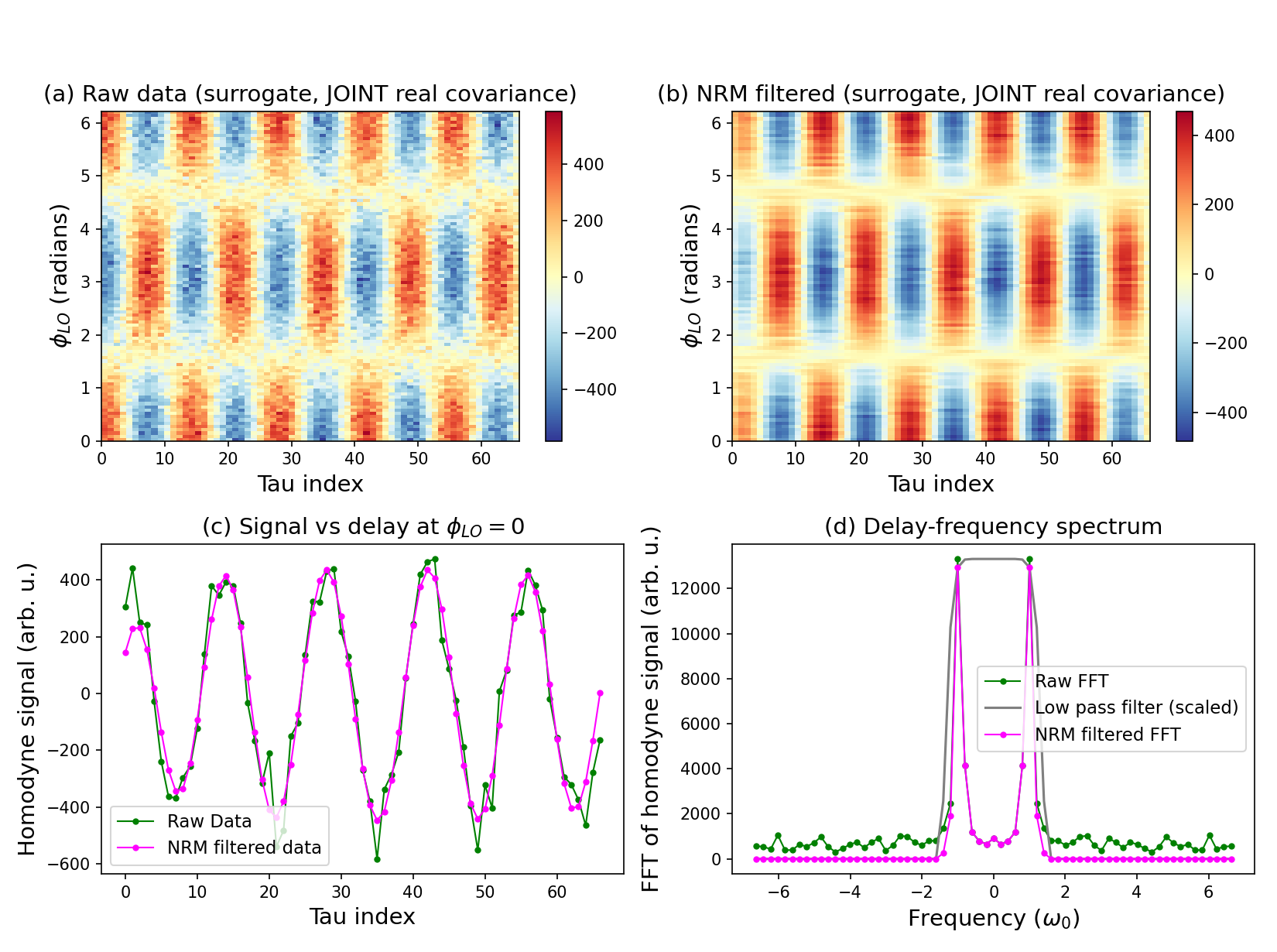}
    \caption{The surrogate data is drawn from a Gaussian state with the same covariance structure as the experimentally measured data, confirming that the NRM filter does not introduce non-Gaussianity.}
\end{figure}\
\end{center}

The classical noise added to the surrogate has two distinct pieces, drawn directly from the physics. The first represents the laser's own brightness fluctuating a little from shot to shot. Since the local oscillator field goes as the square root of its intensity, a small fractional intensity fluctuation $r$ turns into a field change of $(1+r/2)$, using the binomial expansion $\sqrt{1+r}\approx 1+r/2$ for small $r$. Because the measured quadrature is proportional to this field, the same $r$ multiplies the true signal rather than adding to it. We did not attempt to match its exact size to the camera's measured gain mismatch, only its scale. The second piece is independent of the first: one extra offset per $(\tau,\phi_{LO})$ setting, given equally to every one of the $1000$ shots taken at that setting. Because every shot at one setting shares the same offset, taking more shots there can never average it away, unlike ordinary shot noise. 

\begin{center}    
\begin{figure}[h]
    \includegraphics[width=\linewidth]{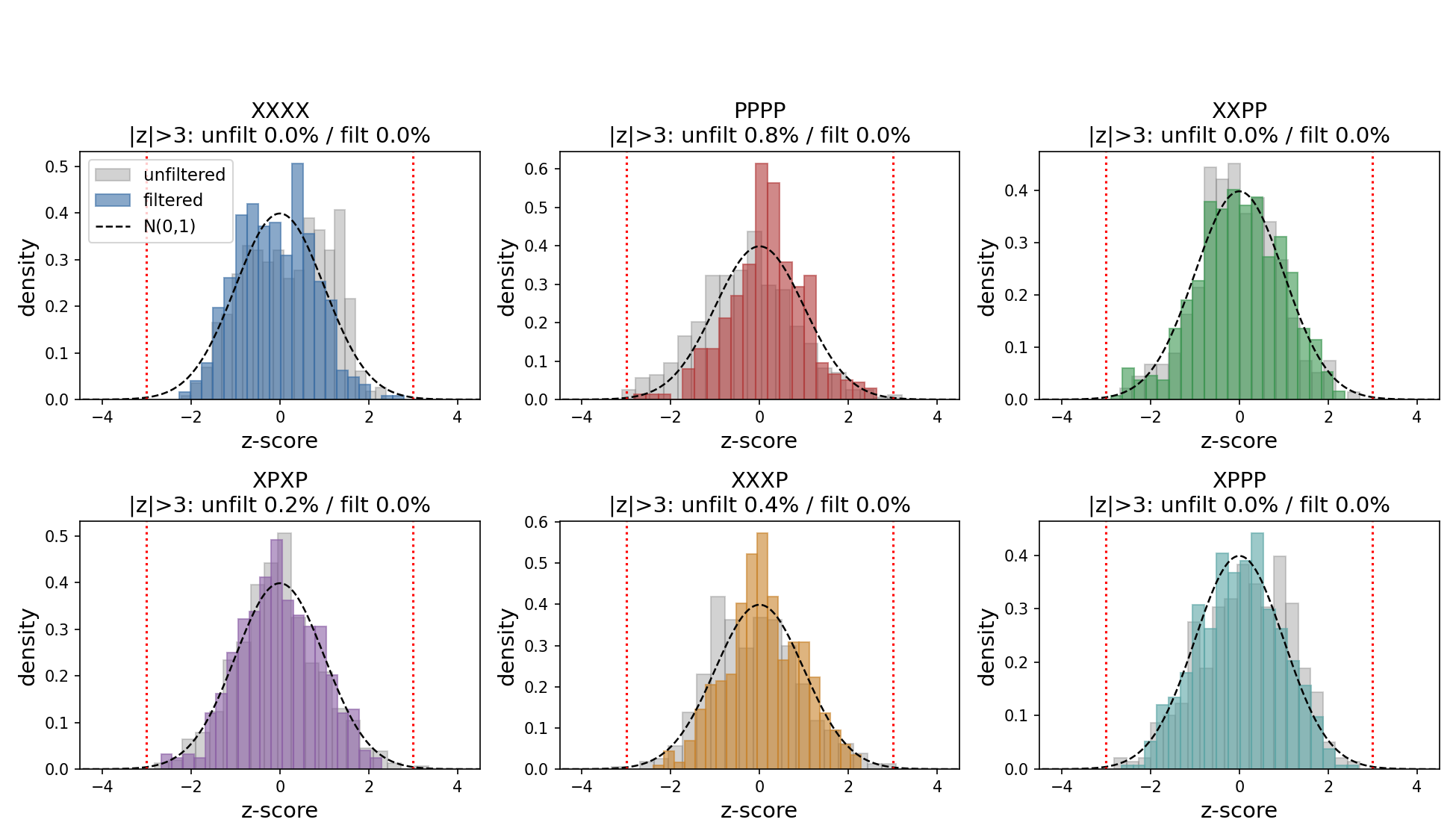}
    \caption{Distribution of z-scores for different operator combinations. Gray bars represent the unfiltered data and colored bars represent the NRM filtered data. The dashed black curve is a standard normal $\mathcal{N}(0,1)$ for reference.}
\end{figure}
\end{center}

After we generate the data with classical noise, the same gain-mismatch and laser-brightness fluctuation, $\epsilon$ and  $\sigma_{\mathrm{RIN}}$, introduced above, we apply the NRM filter. Then, by bootstrapping, we calculate the $z$ score for both filtered and unfiltered data for different operator combinations. For every combination, almost none of the surrogate points land outside the $|z|>3$ lines, in either the filtered or unfiltered case, and the two look essentially the same as one another. This is exactly the check promised earlier; since the squeezing calculation assumes a Gaussian state from the start, this test confirms that assumption holds up after the same filter the squeezing calculation itself relies on.

\subsection{Visibility Extraction}
The visibility V between the signal and the local oscillator is obtained from the mean homodyne fringe. The mean balanced signal at a single spectral bin varies with the LO phase as $\langle D \rangle = 2\sqrt{LS}V\cos{(\phi_{LO} + \phi_{0})}$, where $L$ and $S$ are the LO and signal intensities in the bin. The fringe amplitude is extracted at the near-coherent reference delay, where the signal is brightest and the fit most robust, from the raw shot-averaged fringe. L is measured from an LO-only acquisition at identical camera settings, and S is the difference between the phase-averaged port sum and L. This yields V = 0.89. The dominant systematic uncertainty is in the determination of S as a small difference between the port-sum and LO-only data; a drift of the LO intensity of $1\%$ between the two records shifts V by approximately 0.1. Slow interferometric drift during the phase scan can only reduce the extracted fringe amplitude, so this value is effectively a lower bound on the optical overlap. At the squeezed-state delay the signal is several times weaker, the small-difference systematic is correspondingly larger, and the extracted fringe amplitude is consistent with unit visibility. Since the mode overlap is set by the beam geometry and does not depend on the sub-cycle delay, the reference-delay value is used for the detection efficiency calculation.

\end{document}